%% file: ms_emulateapj.tex
\documentclass[twocolumn]{aastex61}

\usepackage{xspace}
\usepackage{graphicx}
\usepackage{natbib}
\newcommand\msun{\mathrm{M_\odot}}

\newcommand\nh{N_\mathrm{H}}
\newcommand\cmt{\mathrm{cm^{-2}}}
\newcommand\ecs{\mathrm{erg\,cm^{-2}\,s^{-1}}}
\newcommand\es{\mathrm{erg\,s^{-1}}}

\newcommand\atca{\textsl{ATCA}\xspace}

\newcommand\chandra{\textsl{Chandra}\xspace}

\newcommand\heg{{HEG}\xspace}
\newcommand\hetg{{HETG}\xspace}
\newcommand\hetgs{{HETGS}\xspace}

\newcommand\igr{{IGR~J17591$-$2342}\xspace} 
\newcommand\integral{\textsl{INTEGRAL}\xspace}
\newcommand\ibis{{IBIS}\xspace}
\newcommand\isgri{{ISGRI}\xspace}
\newcommand\jemx{{JEM-X}\xspace}
\newcommand\isis{\texttt{ISIS}\xspace}

\newcommand\meg{{MEG}\xspace}

\newcommand\nicer{\textsl{NICER}\xspace}
\newcommand\nustar{\textsl{NuSTAR}\xspace}

\newcommand\rxte{\textsl{RXTE}\xspace}

\newcommand\swiftf{\textsl{Neil Gehrels Swift}\xspace}
\newcommand\swift{\textsl{Swift}\xspace}
\newcommand\xrt{{XRT}\xspace}

\newcommand\xmm{\textsl{XMM-Newton}\xspace}

\newcommand\vlt{\textsl{VLT}\xspace}
\newcommand\xspec{\texttt{XSPEC}\xspace}

\newcommand\aproxgt{\mathrel{%
     \rlap{\raise 0.511ex \hbox{$>$}}{\lower 0.511ex \hbox{$\sim$}}}}
\newcommand\aproxlt{\mathrel{%
     \rlap{\raise 0.511ex \hbox{$<$}}{\lower 0.511ex \hbox{$\sim$}}}}

\received{2019 January}
\accepted{2019 February}

\submitjournal{\apj}

\shorttitle{\chandra-\hetgs Characterization of \igr} 
\shortauthors{Nowak et al.}

\begin{document}

\title{Chandra-HETGS Characterization of an Outflowing Wind
  in the accreting millisecond pulsar \igr}
\author[0000-0001-6923-1315]{Michael A. Nowak}
\affiliation{Physics Dept., CB 1105, Washington University, One
  Brookings Drive, St. Louis, MO 63130-4899}
\author[0000-0001-5067-0377]{Adamantia Paizis} 
\affiliation{Istituto Nazionale di Astrofisica, INAF-IASF, Via Alfonso Corti 12, I-20133 Milano, Italy}
\author[0000-0002-6789-2723]{Gaurava Kumar Jaisawal}
\affiliation{National Space Institute, Technical University of
  Denmark, Elektrovej 327-328, DK-2800 Lyngby, Denmark}
\author[0000-0002-4397-8370]{J\'er\^ome Chenevez}
\affiliation{National Space Institute, Technical University of
  Denmark, Elektrovej 327-328, DK-2800 Lyngby, Denmark}
\author[0000-0002-5769-8601]{Sylvain Chaty}
\affiliation{AIM, CEA, CNRS, Universit\'e Paris-Saclay, Universit\'e 
Paris-Diderot, Sorbonne Paris Cit\'e, F-91191 Gif sur Yvette, France}

\author[0000-0003-3642-2267]{Francis Fortin}
\affiliation{AIM, CEA, CNRS, Universit\'e Paris-Saclay, Universit\'e 
Paris-Diderot, Sorbonne Paris Cit\'e, F-91191 Gif sur Yvette, France}
\author[0000-0002-4151-4468]{J\'er\^ome Rodriguez}
\affiliation{AIM, CEA, CNRS, Universit\'e Paris-Saclay, Universit\'e 
Paris-Diderot, Sorbonne Paris Cit\'e, F-91191 Gif sur Yvette, France}
\author[0000-0003-2065-5410]{J\"orn Wilms}
\affiliation{Dr.\ Karl Remeis-Observatory \& ECAP, University
  of Erlangen-Nuremberg, Sternwartstr. 7, 96049 Bamberg, Germany}
\email{mnowak@physics.wustl.edu, adamantia.paizis@inaf.it, gaurava@space.dtu.dk,
  jerome@space.dtu.dk, chaty@cea.fr, francis.fortin@cea.fr, jrodriguez@cea.fr,
  joern.wilms@sternwarte.uni-erlangen.de}

\begin{abstract}
\igr is an accreting millisecond X-ray pulsar discovered in 2018
August in scans of the Galactic bulge and center by the \integral
X-ray and gamma-ray observatory.  It exhibited an unusual outburst
profile with multiple peaks in the X-ray, as observed by several X-ray
satellites over three months.  Here we present observations of this
source performed in the X-ray/gamma-ray and near infrared domains, and
focus on a simultaneous observation performed with the \chandra-High
Energy Transmission Gratings Spectrometer (\hetgs) and the Neutron
Star Interior Composition Explorer (\nicer).  \hetgs provides high
resolution spectra of the Si-edge region, which yield clues as to the
source's distance and reveal evidence (at 99.999\% significance) of an
outflow with a velocity of $\mathrm{2\,800\,km\,s^{-1}}$.  We
demonstrate good agreement between the \nicer and \hetgs continua,
provided that one properly accounts for the differing manners in which
these instruments view the dust scattering halo in the source's
foreground.  Unusually, we find a possible set of Ca lines in the
\hetgs spectra (with significances ranging from 97.0\% to 99.7\%).  We
hypothesize that \igr is a neutron star low mass X-ray binary at a
distance of the Galactic bulge or beyond that may have formed from the
collapse of a white dwarf system in a rare, calcium rich Type Ib
supernova explosion.
\end{abstract}


\setcounter{footnote}{0}

\section{Introduction}\label{sec:intro}

Accreting msec X-ray pulsars (AMXPs) are a peculiar subclass of
Neutron Star (NS) Low Mass X-ray Binaries (LMXBs). In general, no
X-ray pulsations are detected in classical NS LMXBs: the magnetic
field of the NS is believed to be too faint (${\aproxlt}10^{9}$\,G) to
channel the accreting matter --- provided by the low mass companion
star --- and it ends up being buried in the accretion flow, producing
a `spot-less' accretion on the NS surface. In some cases, however, an
X-ray pulsation is detected: it can be hundreds of seconds (e.g.,
$\sim$120\,s, spinning down to $\sim$180\,s over 40 years in the case
of GX~1+4; see \citealt{jaisawal:18a}) down to the millisecond domain,
in the range of 1.7--9.5\,ms \cite[e.g.,][]{patruno:12a, campana:18a},
the so-called accreting msec X-ray pulsars. Currently 21 such systems
are known \citep{campana:18a}. The fast pulsations are believed to be
the result of long-lasting mass transfer from an evolved companion via
Roche lobe overflow, resulting in a spin up of the NS \citep[the
  recycling scenario;][]{alpar:82a}.  These sources are very important
because they provide the evolutionary link between accreting LMXBs and
the rotation powered millisecond radio pulsars (MSP). Indeed, such a
link has been recently observed in a few systems where a transition
from the radio MSP phase (rotation powered) to the X-ray AMXP phase
(accretion powered) has been detected \cite[transitional MSP;][and
  references therein]{papitto:16a}.

\igr was discovered by \integral during monitoring observations of the
Galactic Centre (PI J. Wilms) and bulge (PI
E. Kuulkers\footnote{http://integral.esac.esa.int/BULGE/}). The source
was detected in a 20--40\,keV \ibis/\isgri (15\,keV -- 1\,MeV;
\citealt{lebrun:03a}) mosaic image spanning 2018 August 10--11 (MJD
58340--58341) at a significance of approximately 9\,$\sigma$ with a
positional uncertainty of 3\,arcmin \citep{ducci:18a}.  \igr was
located at the rim of the field of view of the co-aligned, smaller
field of view instrument \jemx (3--35\,keV, \citealt{lund:03a}), and
hence was not detected in this lower energy bandpass field.

Subsequent observations with the \swiftf satellite refined the
position to within a 3.6\arcsec\ uncertainty (90\% confidence level)
and gave a preliminary estimate of the source's absorbed 0.3--10\,keV
X-ray flux of $(2.6\pm0.2)\times10^{-10}\,\ecs$ \citep{bozzo:18a}.
The spectrum was consistent with a highly absorbed powerlaw ($\nh =
(4.2\pm0.8) \times 10^{22}\,\cmt$, $\Gamma = 1.7 \pm 0.3$).  Optical
follow up observations did not definitively reveal any counterpart
within the \swift error circle \citep{russell:18b}; however, radio
observations showed a counterpart with a flux of $1.09\pm0.02$\,mJy at
$\alpha_\mathrm{J2000.0}=17:59:02.86$,
$\delta_\mathrm{J2000.0}=-23:43:08.0$ (0.6\arcsec\ uncertainty), which
is within 2.1\arcsec\ of the \swift position \citep{russell:18c}.

Joint Neutron Star Interior Composition Explorer (\nicer)/\nustar
observations demonstrated that this radio and X-ray source was in fact
an accreting millisecond X-ray pulsar with a spin frequency of 527\,Hz
and an 8.8\,hr orbital period with a likely companion mass
$>0.42\,\msun$ \citep{ferrigno:18a,sanna:18a}.  The pulsar spin was
detected in both instruments.  The 3--30\,keV \nustar absorbed flux
was $4.2\times10^{-10}\,\ecs$ and the extrapolated 0.1--10\,keV \nicer
flux was $1.3\times10^{-10}\,\ecs$. It was further noted that the
radio flux reported by \citet{russell:18c} was approximately three
times larger than for other observed AMXP (cf.\ \citealt{tudor:17a}).
Near-infrared (NIR) observations with the High Acuity Wide-field
K-Band Imager (HAWK-I) on the \textsl{Very Large Telescope} (\vlt)
further refined the position of \igr to
$\alpha_\mathrm{J2000.0}=17:59:02.87$,
$\delta_\mathrm{J2000.0}=-23:43:08.2$ (0.03\arcsec\ uncertainty;
\citealt{shaw:18a}).  The source was found to be faint in the NIR
($\mathrm{H=19.56\pm0.07}$\,mag and $\mathrm{Ks=18.37\pm0.07}$\,mag,
\citealt{shaw:18a}; see also \S\ref{sec:multi} below).

Starting at 2018 August 23, UTC 17:53 (MJD 58353.74), we used the
\chandra X-ray observatory \citep{weisskopf:02a} to perform a 20\,ks
long Target of Opportunity observation of \igr employing the High
Energy Transmission Gratings Spectrometer (\hetgs;
\citealt{canizares:05a}). As we previously have reported
(\citealt{nowak:18a}, and see \S\ref{sec:chandra} below), our best
determined position for \igr is $\alpha_\mathrm{J2000.0}=17:59:02.83$,
$\delta_\mathrm{J2000.0}=-23:43:08.0$ (0.6\arcsec\ accuracy, 90\%
confidence limit).
This position is consistent with the radio, NIR, and \swift determined
positions (see Figure~\ref{fig:fc}, \S\ref{sec:multi} below).

Bracketing the times of our \chandra observation, proprietary deep
\integral Target of Opportunity observations (August 17--19 and August
25--27, MJD 58347--58349 and 58355--58357, PI Tsygankov;
\citealt{kuiper:18a}) significantly detected the source up to
150\,keV.  The source exhibited a powerlaw spectrum with
$\Gamma=1.92\pm0.05$ in the second observation period, and its 1.9\,ms
pulsation was detected in the 20--150\,keV band at 5.2\,$\sigma$
\citep{kuiper:18a}.

\begin{figure}
\begin{center}
\includegraphics[width=0.29\textwidth,angle=270]{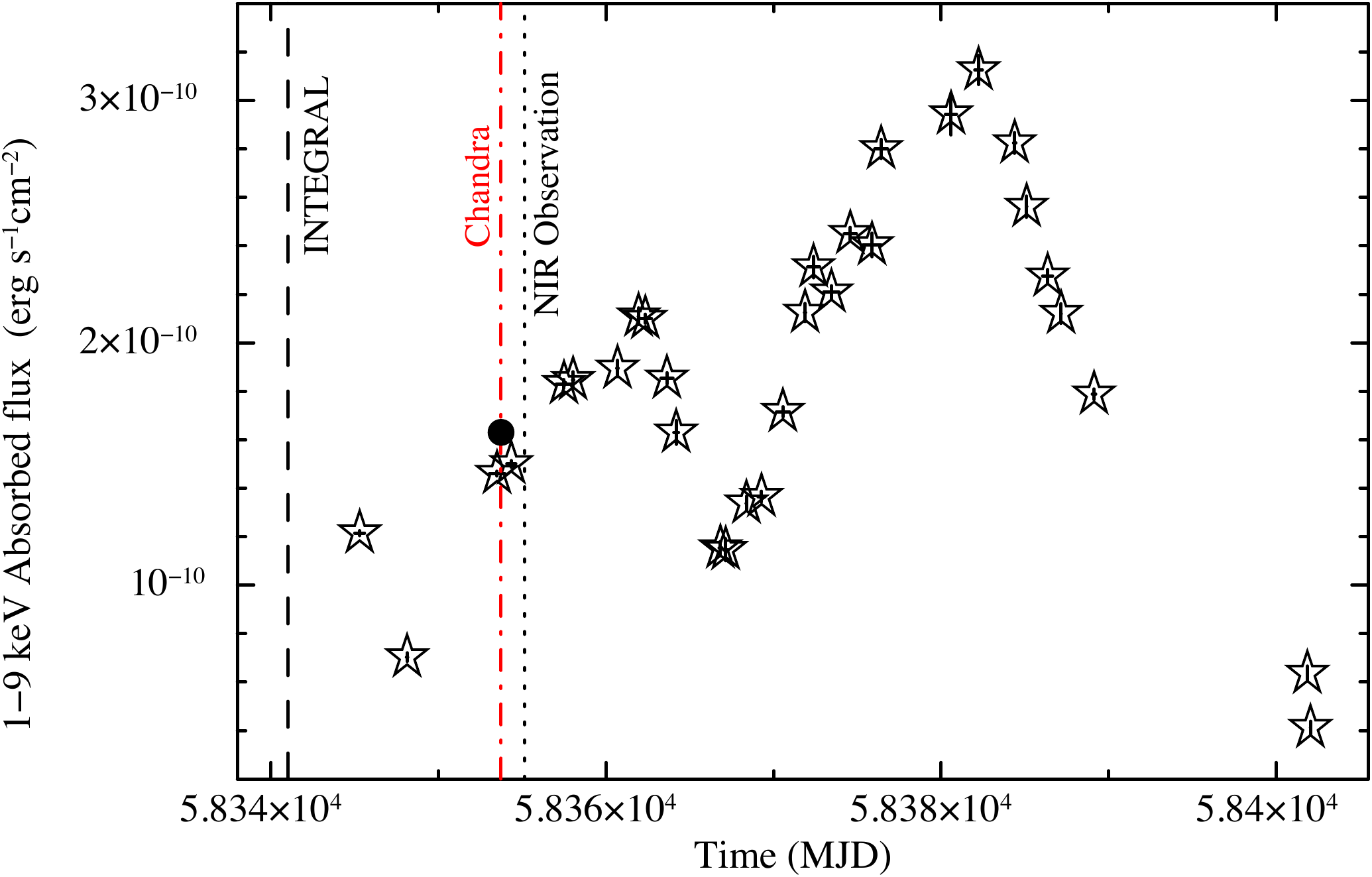}
\end{center}
\caption{Lightcurve showing the absorbed 1--9\,keV flux as determined
  by \nicer observations (observation IDs 1200310101--1200310137). 
  Times of the \integral discovery and our \chandra and NIR observations 
  are also highlighted.} \label{fig:lightcurve}
\end{figure}

An examination of archival \swiftf data showed that the initial
brightening of \igr occurred as early as 2018 July 22 (MJD 58321) and
peaked on 2018 July 25, predating the \integral discovery
\citep{krimm:18a}.  Further \integral observations post initial
discovery showed a rebrightening of the source on 2018 August 30--31
\citep{sanchez:18a,kuiper:18a}.  In Figure~\ref{fig:lightcurve} we
show the \igr lightcurve for the absorbed 1-9\,keV flux as determined
by our analyses of \nicer \citep{gendreau:16a} observations
(observation IDs 12000310101--1200310137; see \S\ref{sec:nicer_obs}
and \S\ref{sec:nicer} below).  The two peaks shown in
Figure~\ref{fig:lightcurve} occur past \emph{at least one} earlier
peak in the lightcurve \citep{krimm:18a}, indicating a complex
lightcurve.  (The degree to which there is further substructure in the
lightcurve over the July/August time frame is difficult to assess,
owing to the disparate bandpasses of the various instruments with
which \igr was observed.)

\igr is the $\mathrm{22^{nd}}$ member of the AMXP class.  A high
resolution X-ray spectroscopic characterization of this system and its
surrounding matter may yield insights as to the evolution of
millisecond pulsars from their accreting low-mass X-ray binary
progenitors.  In this paper, we discuss in detail the \chandra-\hetgs
spectra referenced by \citet{nowak:18a}.  We present evidence of an
ionized outflow with velocity of order 1\% the speed of light, and
attempt to discern local and interstellar absorption.  Taking the
\nicer observations used to create the lightcurve in
Figure~\ref{fig:lightcurve}, we model the spectra that were strictly
simultaneous with our \chandra observation, and discuss the
differences in the model fits that are related to the different fields
of view of these two instruments.  We use \integral observations to
discuss the spectra of \igr above 10\,keV.  We also present new NIR
observations, and discuss their implications.

\section{Observations}\label{sec:obs}

Here we describe observations of \igr performed in several different
energy bands with a variety of instruments.  Although the primary
discovery was obtained by \integral (\S\ref{sec:intro}), our main
focus will be observations obtained with \chandra
(\S\ref{sec:chandra}) and the \textsl{Neutron star Interior
  Composition Explorer} (\nicer; \S\ref{sec:nicer_obs}) observatories.
After describing the observations performed with \chandra and \nicer,
we briefly describe observations obtained with \integral
(\S\ref{sec:integral}) and discuss our follow up optical and IR
observations (\S\ref{sec:multi}).

\subsection{\chandra-\hetgs Observations}\label{sec:chandra}

The \hetgs consists of two sets of gratings, the High Energy Gratings
(\heg, with bandpass $\approx 0.7$--9\,keV and spectral resolution
$E/\Delta E \approx 1\,300$ at 1\,keV) and the Medium Energy Gratings
(\meg, with bandpass $\approx 0.5$--8\,keV and spectral resolution
$E/\Delta E \approx 700$ at 1\,keV), each of which disperses spectra
into positive and negative orders.  Here we consider only $\pm1^{\rm
  st}$ order spectra of the \heg and \meg.  There are too few counts
to produce usable spectra from the higher spectral orders, while the
undispersed $\mathrm{0^{th}}$ order spectra suffers from pileup.  The
first order spectra do not suffer from pileup, as the \emph{peak}
pileup fraction (in \meg $-1$ order near 3.8\,keV where the count rate
peaks at $\approx 0.13\,\mathrm{cts\,s^{-1}}\,\mbox{\AA}^{-1}$) is
$\aproxlt 0.5$\%, and is significantly less for most other orders and
wavelengths. (See \citealt{hanke:09a}.)

Our 20\,ks \chandra data were processed using the suite of analysis
scripts available as part of the Transmission Gratings Catalog
\citep[TGcat;][]{huenemoerder:11a}, running tools from \texttt{CIAO
  v.\,4.10} utilizing Chandra Calibration Database (CALDB) v.\,4.7.8.
The location of the center of the point source's $0^{\rm th}$ order
image was determined by intersecting the dispersion arms via the
\texttt{findzo} tool. This is the position reported by
\citet{nowak:18a}.  Its $0.6''$ accuracy (90\% confidence) is that of
the \chandra aspect solution when no other sources are within the
field of view to further refine the astrometry.

Events within a 16\,pixel radius of the above position were were
assigned to $0^{\rm th}$ order.  This position also defined the
location of the dispersed \heg and \meg spectra.  Any events that fell
within $\pm 16$ pixels of the cross dispersion direction of either the
\heg or \meg spectra were assigned to that grating arm using the
\texttt{tg\_create\_mask} tool.  Spectra were then created with events
that fell within $\pm3$ pixels of the cross-dispersion direction of
the \heg and \meg arm locations (\texttt{tg\_extract}), and assigned
to a given spectral order with \texttt{tg\_resolve\_events} using the
default settings.  Spectral response matrices were created with the
standard tools (\texttt{fullgarf} and \texttt{mkgrmf}).

\subsection{\nicer Observations} \label{sec:nicer_obs}

A series of observations with \nicer were performed throughout the
outburst of \igr (see \citealt{sanna:18a} and
Figure~\ref{fig:lightcurve}). A total of 2.64\,ks were strictly
simultaneous with our 20\,ks \chandra-\hetgs observations. We consider
only these data for purposes of spectral fitting.  There are more
\nicer pointings, likely having a very similar spectral shape and
flux, from periods shortly before or after the datasets that we
consider. Our spectra, however, are already near the limits of the
current understanding of systematic uncertainties in the \nicer
instrumental responses, and therefore inclusion of additional data
would not improve our understanding of the \nicer spectra.

The spectra were extracted using the \nicer tools available in the
\texttt{Heasoft v6.25} package, using calibration products current as
of the release of 2018 November 5.  The response files were
\texttt{ni\_xrcall\_onaxis\_v1.02.arf} and \texttt{nicer\_v1.02.rmf},
which we obtained directly from the \nicer instrument team.  We
created a background file from \nicer observations (with 66\,ks of
effective exposure) of a blank sky field previously observed by the
\texttt{Rossi X-ray Timing Explorer}\footnote{Field number six of
  eight blank sky fields that were previously used for \rxte
  calibration \citep{jahoda:06a}.}  (\rxte).  In all of the fits
described below, rather than include a scaled version of these data as
part of the spectral fit (i.e., ``background subtraction'', even
though what one is doing in ISIS is essentially adding these scaled
background data to the model fits and comparing to the total observed
data), we model the background spectra with a power law (with energy
index $\Gamma \approx 0.78\pm0.02$) and a broad gaussian feature
centered on an energy of $(1.73\pm0.07)$\,keV with width $\sigma
\approx (0.3\pm0.1)$\,keV.  This model is fit to the background data,
while simultaneously incorporating it into the source data model
(without folding it through the spectral response, and appropriately
scaling it for the $\approx 0.04$ relative exposure time of the source
and background).

We bin the \nicer spectra by a minimum of three spectral channels
between 1--6\,keV and four spectral channels between 6--9\,keV.  This
approach ensures that our binning is approximately half width half
maximum of the \nicer spectral resolution (as determined from
empirical measurements of delta functions forward folded through the
\nicer spectral response).  Furthermore, we also impose a
signal-to-noise minimum of 4; however, this latter criterion only
affects the binning of the last few channels in the 8--9\,keV range.

\subsection{\integral Observations} \label{sec:integral}

In order to better understand the long-term behavior of \igr at
energies above 10\,keV, we use \integral
\citep{winkler:03a,winkler:11a} to study its high energy spectra.  We
analyzed all the available \ibis/\isgri data of the monitoring
observations\footnote{Galactic Centre (PI. J. Wilms) and bulge (PI
  E. Kuulkers)} starting from 2018 July 1 (MJD 58300; i.e., prior to
the 2018 July 22 detection with archival data by \citealt{krimm:18a})
up to 2018 October 23 (MJD 58414). Pointings (``science windows'' in
\integral parlance) that had the source within the \ibis/\isgri field
of view ($<$15$^{\circ}$) and with integrated good times $>$1\,000\,s
were used. This resulted in a total of 468 pointings of about 1\,ks
each (none of which were strictly simultaneous with our
\chandra~data). We analyzed the data using the Off-line Scientific
Analysis (OSA) version 11 and the latest instrument characteristic
files (2018 November).

\begin{figure}
\begin{center}
\includegraphics[width=0.40\textwidth,viewport=75 320 559 700,clip]{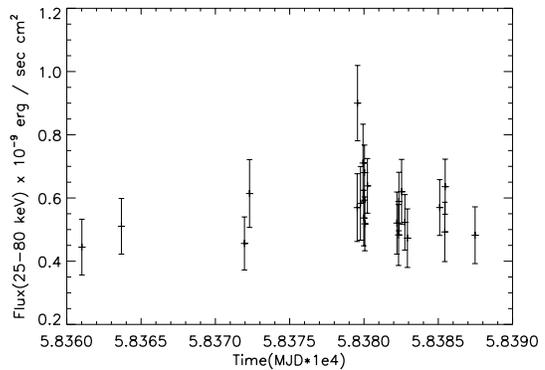}
\end{center}
\caption{\ibis/\isgri 25--80\,keV lightcurve of the hard X-ray
  brightest part of the outburst of \igr (single pointing
  detections).} \label{fig:integral_lcr}
\end{figure}

\igr was detected in 22 pointings in the 25--80\,keV band. In all
these pointings, the source was within ${\sim}9^{\circ}$ from the
center of the field of view. Figure~\ref{fig:integral_lcr} shows the
lightcurve when the source is detected at a pointing level. As can be
seen with respect to Figure~\ref{fig:lightcurve}, the detections
overlap with the highest intensity periods from the \nicer
observations. The 25--80\,keV source flux was obtained using a
$\Gamma=2$ powerlaw spectrum (see \S\ref{sec:int_fit}).

\subsection{Optical/IR Observations} \label{sec:multi}

For optical followup, we triggered optical to infrared observations of
\igr at the European Organisation for Astronomical Research in the
Southern Hemisphere (ESO) of, using the \vlt X-shooter instrument, a
large-band UVB to NIR spectrograph, mounted on the UT2 Cassegrain
focus \citep{vernet:11a}.

\begin{figure}
\begin{center}
\includegraphics[width=0.47\textwidth]{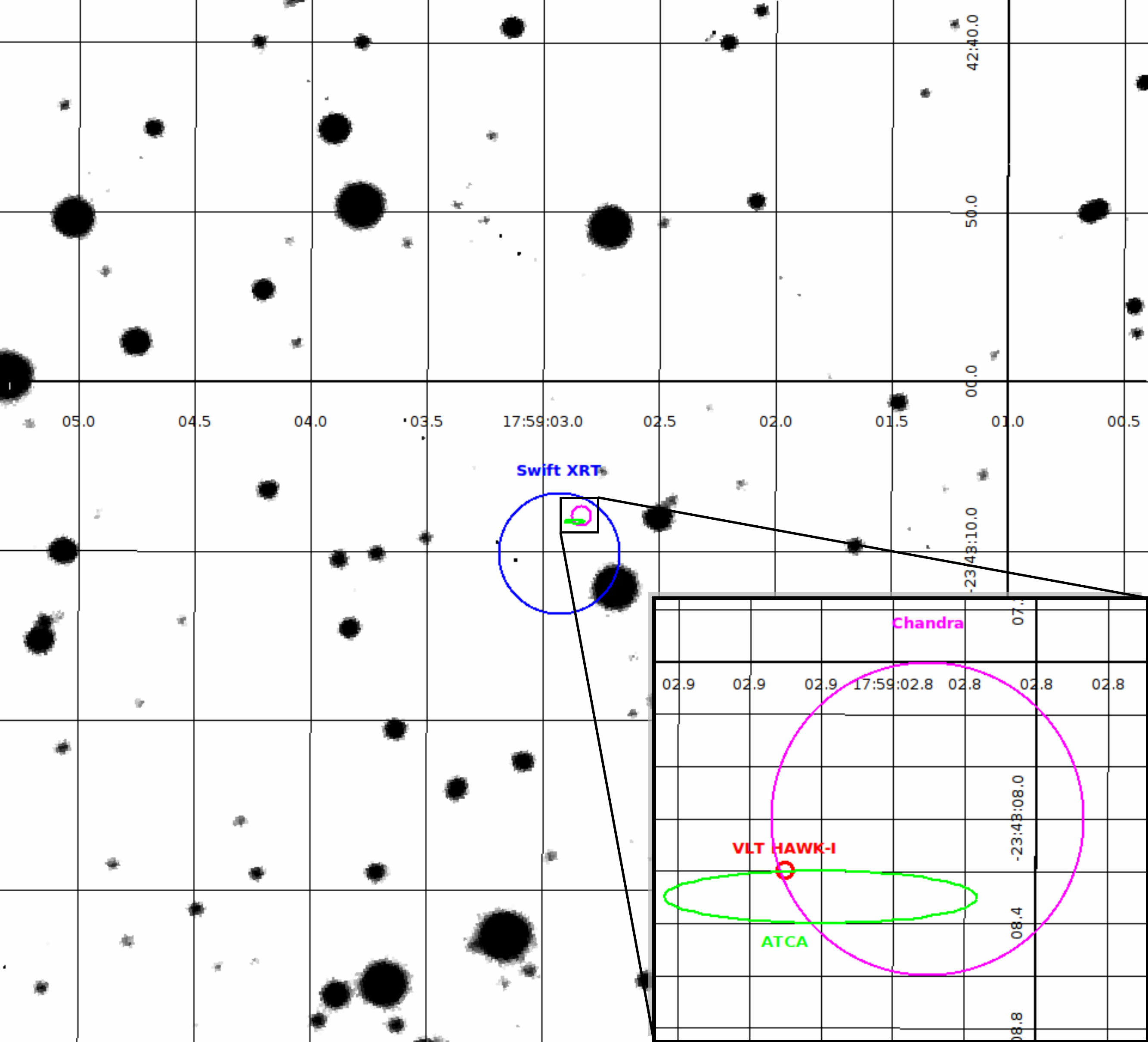}
\end{center}
\caption{The finding chart of \igr in the
I-band, as acquired by X-shooter, indicating the \swift-\xrt,
\chandra, \atca, and \vlt/HAWK-I localization circles.}
 \label{fig:fc}
\end{figure}

We obtained an I-band acquisition image on 25 August 2018, UTC 03h30
(exposure time 120\,s). Figure~\ref{fig:fc} shows this finding chart
of \igr in the I-band, as acquired by X-shooter, indicating the
\swift-\xrt, \chandra, \atca, and \vlt/HAWK-I localization circles
referenced in \S\ref{sec:intro}. From the image we derive a lower
limit for the I-magnitude of the counterpart of the X-ray source of
$\mathrm{I}\ge 24.7\pm0.6$\,mag (Johnson filter, magnitude of the
source at $3\,\sigma$ above the sky noise).  The uncertainty on the
determined magnitude is rather large because we used a mean zero-point
to flux-calibrate the photometry. This value in I-band is consistent
with the H and Ks values obtained with HAWK-I observations
\citep{shaw:18a}.

We also obtained NIR spectra on 25 August 2018, UT03h33 to UT04h37
(exposure time of 64\,m, with airmass between 1.346--1.836) that we
analyzed by performing a standard reduction using the
\texttt{esoreflex} pipeline \citep{freudling:13a}.  We detect a very
faint spectrum ($\mathrm{S}/\mathrm{N}\sim3.2$ at the maximum of the
whole band coverage), as expected for a faint source with the I, H,
and Ks band values discussed above.  By flux-calibrating the faint
spectrum, we find $\mathrm{F_\nu < 0.12}$\,mJy at the K-band
wavelength of 2.2\,$\mu$m (i.e., corresponding to K$> 16.8$\,mag). By
applying a median filter we detect a faint continuum signal at the
level of $\mathrm{F_\nu = 0.025}$\,mJy (i.e., K$= 18.6$). Both
measurements are consistent with the Ks value obtained with HAWK-I
observation.

We point out here that in absence of detection of variability of the
NIR candidate counterpart, we can not unambiguously associate either
the candidate counterpart claimed by \citet{shaw:18a}, nor the faint
spectrum we detected with X-shooter, to the variable X-ray source.

For the value of the equivalent neutral absorption column, $\nh =
(4.9\pm0.2) \times 10^{22}\,\cmt$, that we originally reported
(\citealt{nowak:18a}; where we used the absorption model, cross
sections, interstellar medium abundances discussed by
\citealt{wilms:00a}), the corresponding V-band absorption is
$\mathrm{Av \approx24.6\,mag}$ (using the relationship between
equivalent neutral column and V-band absorption given by
\citealt{predehl:95a}; although see our more detailed discussions of
absorption modeling below) and the K-band absorption is
$\mathrm{Ak\approx2.77}$\,mag (using \citealt{fitzpatrick:99a}).

The I$-$K color value being greater than at least 6\,mag suggests a
late spectral type companion star, located at the distance of the
Galactic bulge.

\section{Spectral Fits}\label{sec:specfit}

\subsection{Hard X-ray Continuum Fits}\label{sec:int_fit}

\begin{figure}
\begin{center}
\includegraphics[width=0.3\textwidth,angle=270,viewport=75 25 550 705,clip]{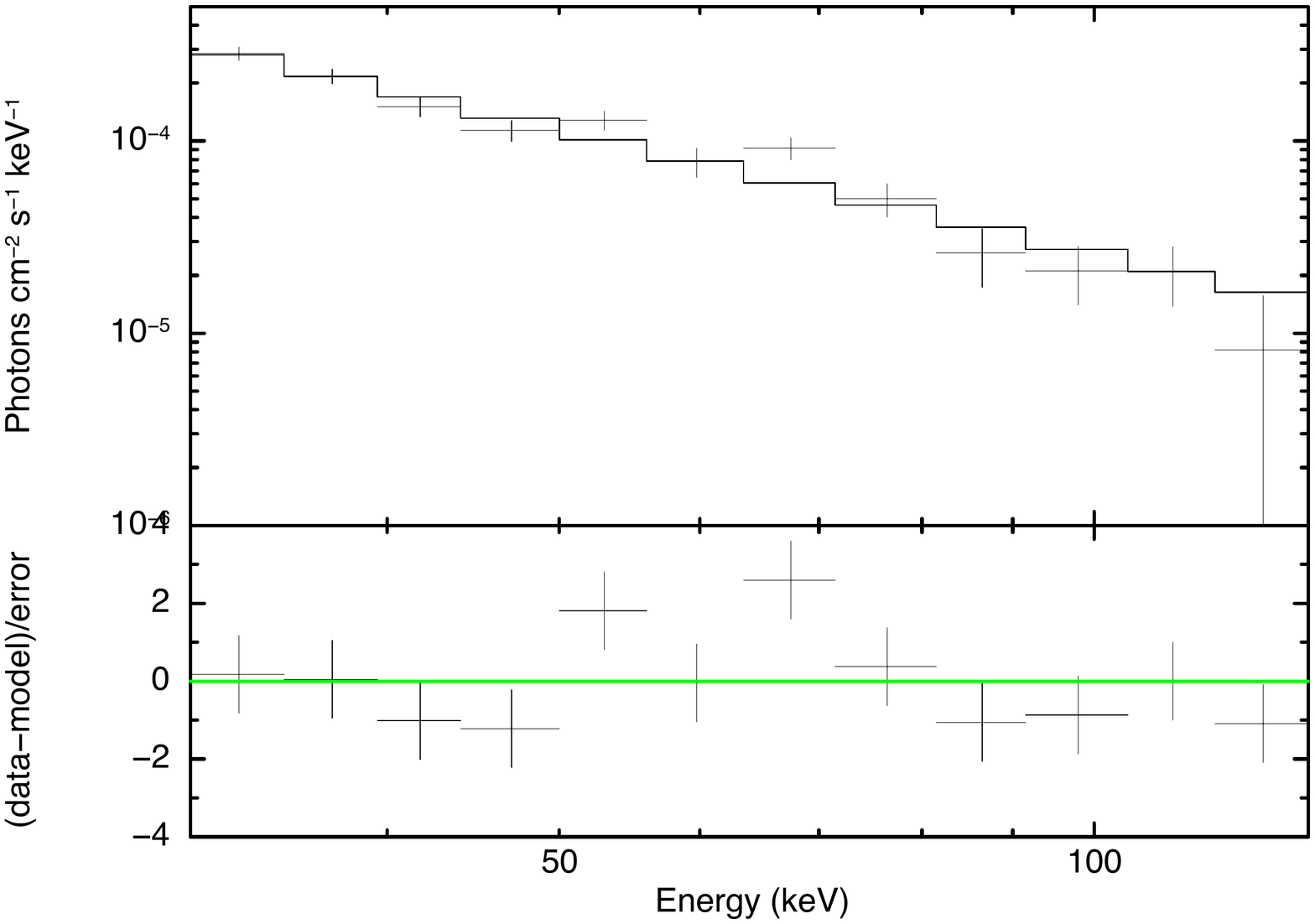}
\includegraphics[width=0.32\textwidth,angle=270,viewport=83 30 542 711,clip]{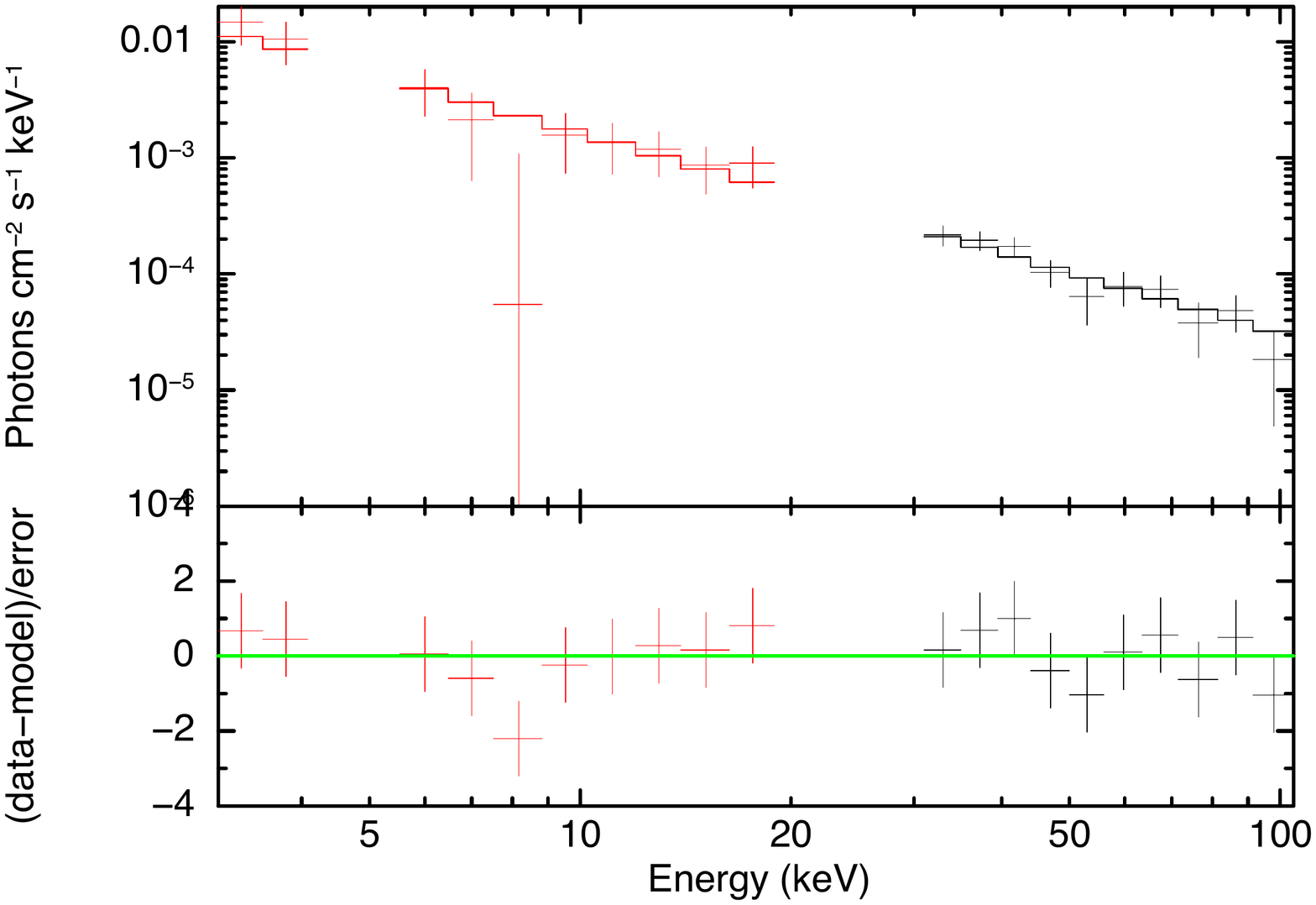}
\end{center}
\caption{\integral~spectra of \igr. \textit{Top panel}: \ibis/\isgri
  average spectrum and best fit of the data points shown in
  Figure~\ref{fig:integral_lcr} (effective exposure
  $\sim$24\,ks). \textit{Bottom panel}: simultaneous \jemx (red) and
  \ibis/\isgri (black) spectrum and best fit of five pointings with
  the source within the \jemx fully coded field of view (effective
  exposure $\sim$6\,ks with \ibis/\isgri and $\sim$8\,ks with
  \jemx). See text.} \label{fig:integral_spe}
\end{figure}

We first consider the \ibis/\isgri spectra obtained from the average
of the 22 pointings discussed above (i.e., the observations
represented in the lightcurve shown in Figure~\ref{fig:integral_lcr}).
We fit these spectra with \xspec v12.9.1 \citep{arnaud:96a} using a
powerlaw, and found a spectrum with $\Gamma$=$2.0 \pm 0.2$ (reduced
$\chi^2=1.29$, 10 degrees of freedom). These \integral observations
extend the bandpass beyond the $\approx 70$\,keV upper limit of
\nustar observations, and here we find that \igr is detected up to
about 110\,keV with no improvement obtained with the addition of a
cutoff, even going out to $\approx 200$\,keV
(Figure~\ref{fig:integral_spe}, top panel).

We selected five pointings for which \igr was both bright (second peak
from Figure~\ref{fig:lightcurve}, between MJD 58380--58383) and within
the \jemx fully coded field of view (where the detection significance
is maximum, $<3^{\circ}$). This selection resulted in a sub-sample of
five science windows (ID: 200100250010, 200100340010, 200200250010,
200200330010, 200200340010).  The simultaneous \ibis/\isgri and \jemx
spectra of the five average pointings (Figure~\ref{fig:integral_spe},
bottom panel) resulted in a best fit (reduced $\chi^2=0.66$, 19
degrees of freedom) powerlaw spectrum with $\Gamma=1.8 \pm 0.2$ and
frozen neutral hydrogen N$_{H}=3.3\times10^{22}\,\mathrm{cm}^{-2}$
(taken from model E in Table~\ref{tab:pars}, as discussed below in
\S\ref{sec:continuum}).  Again, no cutoff is required within the
\integral band.  The average absorbed fluxes are
$\mathrm{F_{25-80\,keV}}=4.7\times10^{-10}\,\ecs$,
$\mathrm{F_{3-25\,keV}}= 5.4\times10^{-10}\,\ecs$, and
$\mathrm{F_{1-9\,keV}}=2.8\times10^{-10}\,\ecs$.  These values are
compatible with the \nicer fluxes shown in
Figure~\ref{fig:lightcurve}.  Assuming a distance of 8\,kpc, they
correspond to luminosities of
$\mathrm{L_{25-80\,keV}}=3.6\times10^{36}\,\es$,
$\mathrm{L_{3-25\,keV}}=4.1\times10^{36}\,\es$, and
$\mathrm{L_{1-9\,keV}}=2.1\times10^{36}\,\es$.

A deeper analysis of the \integral data (mosaicking detections,
spectral variability and timing) is beyond the scope of this paper.
  
\subsection{Soft X-ray Continuum and Line Fits}\label{sec:continuum}

All further analyses presented below were performed with the
\texttt{Interactive Spectral Interpretation System} (\isis;
\citealt{houck:00a}).  In order to increase the signal-to-noise ratio
of our spectra, we combine the positive and negative first order \heg
and \meg spectra using the \isis functions
\texttt{match\_dataset\_grids} (to match the \heg wavelength grid to
that of \meg) and \texttt{combine\_datasets}\footnote{The
  \texttt{combine\_datasets} function essentially adds together the
  product of exposure, effective area, and response function for each
  individual observation within the standard forward folding equation
  \citep{davis:01b}, while also properly combining the background.  It
  has been well-vetted via comparisons against standard
  \texttt{Heasoft} and \texttt{CIAO} functions for combining spectral
  responses and backgrounds.}. We limit the energy range to 1--9\,keV,
but do not further bin the data, and use \citet{cash:79a} statistics
in the fits so as to facilitate the spectral line search, without
biasing against absorption lines (see below).

We use a continuum model similar to the one \citet{sanna:18a} employed
to fit joint \nicer/\nustar data of \igr, specifically an absorbed
(\texttt{tbvarabs}; \citealt{wilms:00a}, where we have also adopted
the atomic cross sections and interstellar medium abundances discussed
in that work) blackbody (\texttt{bbodyrad}) plus Comptonization
(\texttt{nthcomp; \citealt{zdziarski:99b}}) spectrum.  Lacking
simultaneous data above 9\,keV, we do not have good leverage on some
of the Comptonization parameters, so for all models we fix the coronal
temperature to the 22\,keV value found by \citet{sanna:18a} such that
we can more readily compare to their findings.  Our results in the
1--9\,keV band are not sensitive to the coronal temperature; however,
we note that a 22\,keV coronal temperature would imply a spectral
curvature in the 50--150\,keV band that we do not see in the \integral
spectra shown in Figure~\ref{fig:integral_spe}.  The remaining
\texttt{nthcomp} parameters are the normalization ($\mathrm{N_{nc}}$),
the Compton powerlaw photon index ($\mathrm{\Gamma_{nc}}$), and the
seed photon temperature ($\mathrm{kT_{nc}}$).  The latter is tied to
the blackbody temperature.  The remaining \texttt{bbodyrad} parameter
is its normalization, $\mathrm{N_{bb}}$, which nominally corresponds
to $\mathrm{R^2_{km}/D^2_{10\,kpc}}$, where $\mathrm{R_{km}}$ is the
source radius in km, and $\mathrm{D_{10\,kpc}}$ is the source distance
in units of 10\,kpc.

We include one other component in our model, not found in the \nicer
modeling of \citet{sanna:18a}, namely a dust scattering component
using the \texttt{dustscat} model \citep{baganoff:03a}. This component
accounts for the scattering of soft X-rays out of our line of sight
due to dust grains (see the discussion of
\citealt{corrales:16a}). Taking this effect into account is important
for the high spatial resolution measurements done with \chandra, which
resolve \igr into a point source and an arcminute size dust scattering
halo. In contrast, as we further discuss below, the halo emission is
included in the overall \nicer spectrum owing to the arcminute scale
resolution line of sight provided by this instrument. Following
\citet{nowak:12a}, in our \chandra analysis we therefore tie the halo
optical depth to a value of $\tau_{scat} =
0.324\,(\nh/10^{22}\,\cmt)$, where $\nh$ is the equivalent neutral
column obtained from the \texttt{tbvarabs} model.  The dust halo size
relative to the instrumental point spread function (PSF) is frozen at
$\mathrm{H_{size}}=200$ (i.e., nearly all scattered photons are lost).

Our continuum model with the dust scattering halo describes the \hetg
spectra well (Cash statistic = 2048.6 for 2200 degrees of freedom),
with a fitted equivalent neutral hydrogen column of $\nh = (4.4\pm0.2)
\times 10^{22}\,\cmt$.  The modeled 1--9\,keV absorbed flux is
$1.58\times 10^{-10}\,\ecs$. (All implied 1--9\,keV absorbed fluxes
for the models discussed below fall within a few percent of this value
for the \chandra-\hetgs spectra.)  Assuming a distance of 8\,kpc, this
translates to an absorbed, isotropic luminosity in the 1--9\,keV band
of $1.21\times10^{36}\,\ecs$

There are a number of prominent, narrow residuals in the spectra,
especially near the Si edge region.  To assess these residuals, we
perform a ``blind line search'' (see the description of this
functionality in \citealt{nowak:17a}).  We write our model (using
\isis notation) as follows:
\begin{eqnarray}
\mathtt{tbvarabs*powerlaw}&*&\mathtt{exp(lines/bin\_width\_en)} \\
 + 0&*&(\mathtt{constant}+...) ~~, \nonumber
\end{eqnarray}
where $\mathtt{tbvarabs}$ is the \citet{wilms:00a} absorption model,
$\mathtt{powerlaw}$ is the standard function with photon energy index
$\Gamma$, $\mathtt{exp}$ is an exponential function, and
$\mathtt{bin\_width\_en}$ is a function that returns the width of a
data spectral bin in keV.  The function $\mathtt{lines}$ is defined
within the search script and returns a sum of standard
$\mathtt{gaussian}$ fit functions, which as defined in \isis or \xspec
are line profiles integrated within each data bin.  It is for this
latter reason that we divide by the data bin widths, so that any
rebinning of the data will not strongly affect the fit parameters. We
multiply the continuum by line functions within an exponential to
ensure that the model never yields negative counts and so that it can
smoothly pass from absorption to emission lines.  The (multiple)
$\mathtt{constant}$ functions (multiplied by 0 so as not to add to the
continuum) are used as ``dummy parameters'' to allow us to transform
the parameters of the $\mathtt{gaussian}$ line functions.  Rather than
fit a line amplitude, we instead fit a parameter closer to line
equivalent width.  Further, as a line becomes significantly more
absorbed, we increase its equivalent width by increasing the line
width, rather than by increasing the magnitude of the line
amplitude. Since the data are not good enough to distinguish between
being on the damping wings of the equivalent width curve of growth and
a true increase in line width, we find that increasing the line width
is numerically more stable.  We limit all line widths to lie between
values $\sigma=0.1$--20\,eV.

In the line search procedure, we add an additional $\mathtt{gaussian}$
function to the $\mathtt{lines}$ function, and while holding the
continuum and any previously detected lines fixed, we fit the
parameters of this added line feature allowing its amplitude, width,
and energy (within a limited range) to be free parameters. We store
the change in fit statistics and the parameters of the fitted line.
We scan along the full energy range of the spectra in this manner.
The ten features with the largest change in fit statistic are then
individually refit, now with both the continuum parameters and
previous line fits allowed to vary.  The feature leading to the
largest improvement in fit statistic is then added to the model, and
the scan is repeated.  (At this stage, no error bars are determined
for the line fits.)

Results for the eleven most significant features found by this method
are presented in Table~\ref{tab:blind}.  Possible line identifications
are also presented, along with the line redshifts if these
identifications are in fact correct.  These features were found in
fits to the combined spectra; however, we visually inspected the
combined fits applied to the spectra from the individual gratings
arms, as well as applied to the spectra for just the combination of
the \heg spectra and just the combination of the \meg spectra.  The
fitted features were consistent with these individual spectra, albeit
with noisier statistics.  None of the features appeared to be the
result of a single spectrum or a single combination of spectra, as
might be the case for an interloping faint source coincident with one
gratings arm, or an unmodeled response feature limited to a subset of
the arms.

Several significant features are found near the location of the
expected Si absorption K-edge, so we include these in subsequent
models, constraining the line energies to lie within a 10\,eV interval
and to have widths $\sigma<10$\,eV.  The possible blueshifted
\ion{Si}{13} Ly$\alpha$ feature (see \citealt{hell:16a} for the most
recent measurements of its energy) is fairly significant, so we
include it in all subsequent fits, and further add \ion{Si}{13}
$\beta$ and $\gamma$ lines tied to the same blueshift and relative
line width.  Likewise, we include both the Si K$\alpha$ line at
1.7349\,keV (whether this is a real feature, or an unmodeled component
in the \hetg response function) and the 1.848\,keV feature near the Si
edge (see discussion below). Although we have no good identification
for the absorption feature near 1.695\,keV, its presence may affect
our models of the Si edge.  We also include this feature in all
subsequent models.

The possible presence of Ca features is somewhat unusual, but many of
these features are formally more statistically significant than, e.g.,
the possible Si K$\alpha$ absorption line.  Given that they may
provide some information about the nature of an evolved companion, we
include them in all subsequent fits constraining their line energies
to lie within a 50\,eV interval and to have widths $\sigma<20$\,eV.
They do not strongly influence any of the continuum or absorption
parameters, as verified by Markov Chain Monte Carlo error contours for
all models and parameters discussed below.  We do not attempt to tie
these features to a single common Doppler shift.

Lacking any plausible identification for the 1.009\,keV or 3.47\,keV
features, we do not include them in subsequent models.  The former
feature consists of only a few events in a very faint portion of the
observed spectrum (hence its large equivalent width, despite being
only a few detected events).  The latter feature does not strongly
influence the remaining fit parameters.

\begin{deluxetable}{crrrcr}
\setlength{\tabcolsep}{0.03in} \tabletypesize{\footnotesize}
\tablewidth{0pt} \tablecaption{Results of Blind Line Search}
\tablehead{\colhead{$E_{obs}$} & \colhead{$\Delta C$} & \colhead{EW} &
  \colhead{Order} & \colhead{ID} & \colhead{$z$} \\ \colhead{keV} & &
  \colhead{eV} } 
\startdata 
1.0088 & $-12.8$ & $  98$ &  1 & \nodata       & \nodata \\ 
1.6949 & $-9.3 $ & $-2.3$ &  3 & \nodata       & \nodata \\ 
1.7349 & $-7.5 $ & $-2.1$ &  9 & Si K$\alpha$  & 0.003   \\ 
1.8481 & $-8.4 $ & $-3.2$ &  6 & Si near edge  & \nodata \\ 
1.8825 & $-23.3$ & $-4.3$ &  0 & \ion{Si}{13}r & -0.009  \\ 
2.2181 & $-8.4 $ & $-4.2$ &  4 & \ion{Si}{13}b & -0.016  \\ 
3.4727 & $-8.2 $ & $ 5.3$ &  7 & \nodata       & \nodata \\ 
3.6865 & $-9.3 $ & $-5.6$ &  2 & Ca K$\alpha$  & 0.0004  \\ 
3.8447 & $-8.3 $ & $ 6.1$ &  5 & \ion{Ca}{19}i & 0.010   \\ 
3.8963 & $-7.3 $ & $-6.4$ & 10 & \ion{Ca}{19}r & 0.016   \\ 
4.2955 & $-7.9 $ & $ 4.8$ &  8 & \ion{Ca}{20}a & 0.044   \\ 
\enddata 
\tablecomments{Results from a blind search to the
  unbinned, combined \chandra-\hetg spectra, using a model consisting
  of an absorbed/scattered Comptonized spectrum.  The columns give the
  fitted energy of the line, the change in Cash statistic when
  including the line, the line equivalent width (negative values are
  absorption, positive are emission), the order in which the lines were
  added (numbers 0--10), a potential line ID, and an implied redshift
  (negative values for blueshifts) if this ID is
  correct.}\label{tab:blind}
\end{deluxetable}

The above model fits the data well with a Cash statistic of 1963.0 for
2172 degrees of freedom; however, it requires a fitted equivalent
neutral column of $\nh = (5.1^{+0.3}_{-0.1})$ $\times10^{22}\,\cmt$.
This value is somewhat larger than the $\nh =
(3.6\pm1.1)$--$(4.4\pm0.3) \times 10^{22}\,\cmt$ values found by
\citet{sanna:18a} (\swift/\nicer/ \nustar/\integral) and
\citet{russell:18a} (\swift), respectively.  Our higher equivalent
neutral column value is in part driven by the need to describe the
complexities of the Si K-edge region, as well as possibly also by
differences in the abundance sets used.  We consider this region
further in \S\ref{sec:edge} below.

\subsection{Si Edge Region Models}\label{sec:edge}

Recently, \citet{schulz:16a} have published a \chandra-\hetgs study of
the Si K-edge region of Galactic X-ray binaries with (continuum)
fitted equivalent neutral columns in the range of $\nh \approx (1-6)
\times 10^{22}\,\cmt$.  Among the conclusions of this survey are the
following: 1) the absorption model of \citet{wilms:00a}, when using
their adopted interstellar medium (ISM) abundances, \emph{under
  predicts} the depth of the Si K-edge, 2) the edge itself is complex,
3) there often is a near-edge absorption feature at $\approx
1.849$\,keV that even in a single source appears to have a variable
equivalent width that is loosely correlated with fitted $\nh$, and 4)
there often is a \ion{Si}{13} absorption feature also with variable
equivalent width with an even weaker correlation with fitted $\nh$.
The variability of the latter two features indicates that for many of
the eleven X-ray binaries included in the survey of
\citet{schulz:16a}, some fraction of the observed absorption is local
to the system, as opposed to being more broadly distributed throughout
the ISM.

The near edge and \ion{Si}{13} features are already accounted for in
our models.  The energy of the near edge feature is consistent with
the values found by \citet{schulz:16a}, and therefore this feature is
likely ``at rest'' relative to its expected energy.  On the other
hand, we do not find \ion{Si}{13} at rest but instead find a
\emph{blueshifted} velocity of $\mathrm{\approx 2\,800\,km\,s^{-1}}$
with $\mathrm{\sigma \approx 200\,km\,s^{-1}}$.  This is in contrast
to \citet{schulz:16a} who found the magnitude of any red or blueshifts
to be $\mathrm{\aproxlt 200\,km\,s^{-1}}$, but found velocity widths
on the order of $\mathrm{\aproxlt 700\,km\,s^{-1}}$.

\begin{figure}
\begin{center}
\includegraphics[width=0.47\textwidth,viewport=105 40 451
  532,clip]{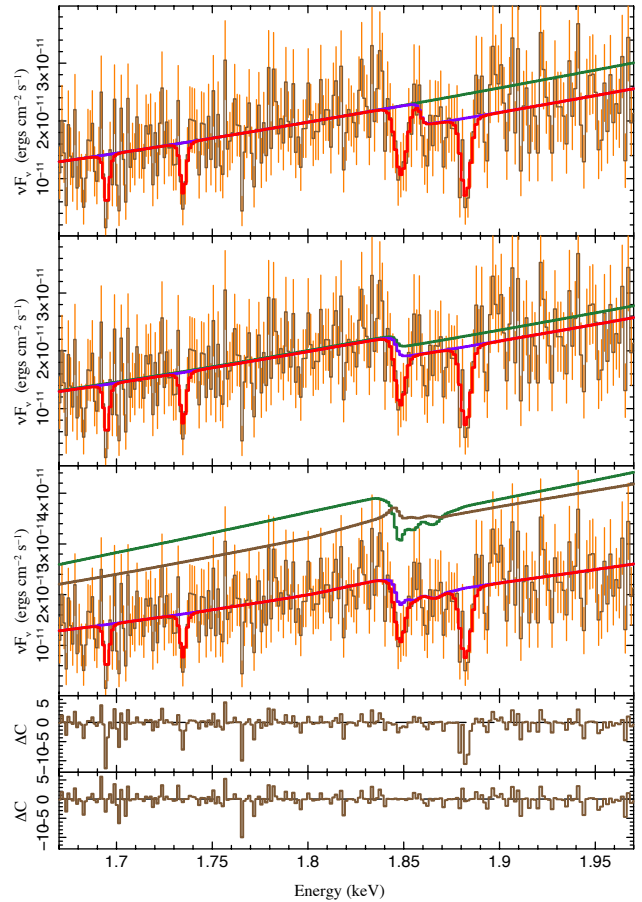}
\end{center}
\caption{The silicon K-edge region for the flux-corrected combined
  \heg and \meg first order \chandra-\hetg spectra (brown with orange
  1\,$\sigma$ error bars), fit with an absorbed Comptonized spectrum
  (see \S\ref{sec:continuum}).  In all panels the red line shows the
  full fitted model, and the purple line shows the model with the
  absorption lines removed.  In the top panel, a phenomenological edge
  has been used to describe the Si K-edge (model A). The green line
  shows the model with edge and lines removed.  In the second panel,
  the Si K-edge absorption has been modeled by allowing a freely
  variable (increased) Si abundance in the ISM (model B). The green
  line shows the model with the Si abundance set to the solar value
  and the lines removed.  The third panel shows the model using the
  dust scattering and absorption models of
  \protect{\citet{corrales:16a}} (model C).  The green line shows the
  model with only the dust absorption contribution.  The brown line
  shows the model with only the dust scattering contributions. The
  fourth panel shows the residuals for model C with the lines removed.
  The fifth panel shows the residuals for model C.}
 \label{fig:edge_phenom}
\end{figure}

To further compare with the phenomenological models of
\citet{schulz:16a}, we take the continuum models of
\S\ref{sec:continuum} and modify the \texttt{tbvarabs} Si edge by
either reducing the Si abundance to 0.01 of the ISM value and adding a
phenomenological \texttt{edge} model (with parameters
$\mathrm{E_{edge}}$ and $\mathrm{\tau_{edge}}$), or by instead
allowing the Si abundance ($\mathrm{A_{Si}}$) to be a free parameter.
These models are referred to as models A and B, respectively, in
Table~\ref{tab:pars}, and the flux corrected spectra\footnote{Flux
  correction is performed on both the model and data counts using the
  \isis \texttt{flux\_cor} function, which only relies on the detector
  response and thus for the case of the detected counts is independent
  of the model.} in the Si edge region are shown in the top two panels
of Figure~\ref{fig:edge_phenom}.  For both models, the equivalent
neutral column is reduced to a value of $\nh \approx 4.2 \times
10^{22}\,\cmt$, with either the abundance increasing to a value of
$\mathrm{A_{Si}=1.76^{+0.49}_{-0.46}}$, or the phenomenological edge
requiring an optical depth of $\mathrm{\tau_{edge} = 0.19\pm0.05}$.
Both of these results are completely consistent with those found by
\citet{schulz:16a}, with the value of the added edge optical depth
relative to fitted equivalent neutral column falling within the data
range shown in their Figure~5. The \citet{schulz:16a} results also
indicate edge optical depth values approximately twice that predicted
by the \citet{wilms:00a} model, consistent with our findings for the
fitted abundance in our model B.  We note, however, that the magnitude
of our fitted value for the equivalent width of the near edge feature
is $\approx 50$\% larger than the largest values found by
\citet{schulz:16a}.  We further discuss this result below.

Also as discussed by \citet{schulz:16a}, the location of the edge has
a degree of uncertainty due to the presence of the near edge
absorption feature.  In fact, our best fit edge energy is higher than
the absorption feature energy, although the error bars allow for the
edge to be at 1.844\,keV which is the expected location for neutral Si
at rest.

We next consider a more physical model for the edge region.
Specifically, we use the dust scattering and edge models of
\citet{corrales:16a}, which in their \isis
implementations\footnote{Available via
  https://github.com/eblur/ismdust/releases.} are broken up into
individual absorption and scattering components for both silicate and
graphite dust grains.  We multiply the scattering components by an
energy-dependent factor in an identical manner as for the
\texttt{dustscat} model \citep{baganoff:03a} to account for the
fraction of flux that scatters back into our line of sight given the
finite size of the instrumental PSF. We again parameterize this factor
with a fixed value of $\mathrm{H_{size}=200}$ (i.e., nearly all
scattered photons are lost from the spectrum).  We further tie the
dust components to the fitted equivalent neutral column via two
parameters: the mass fraction of the ISM column in dust
($\mathrm{f_{dust}})$, and the fraction of dust in silicates
$\mathrm{f_{silicate}}$.  We fit one model where these values are free
parameters (model C), and one model where they are frozen to their
commonly presumed values (see \citealt{corrales:16a}) of
$\mathrm{f_{dust}=0.01}$ and $\mathrm{f_{silicate}=0.6}$ (model D).
Both models fit the data extremely well, as seen in
Figure~\ref{fig:edge_phenom} and Table~\ref{tab:pars}.

\begin{figure}
\begin{center}
\includegraphics[width=0.47\textwidth,viewport=90 48 580 520,clip]{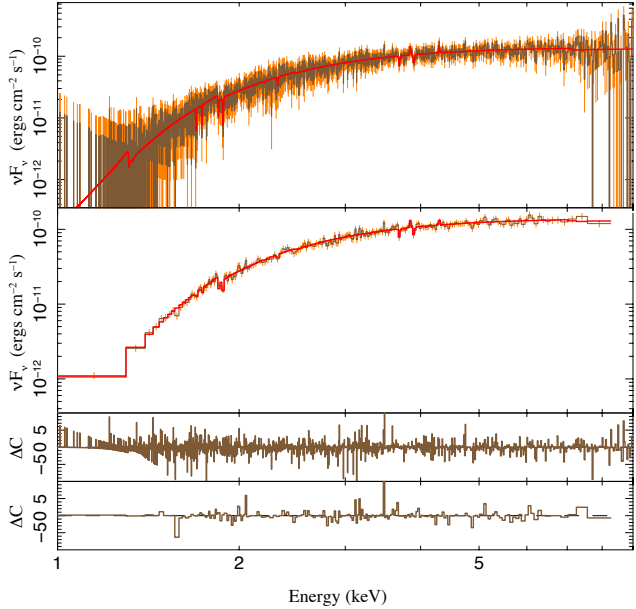}
\end{center}
\caption{The flux-corrected combined \heg and \meg first order
  \chandra-\hetg spectra (brown with orange 1\,$\sigma$ error bars),
  fit in the 1--9\,keV band with an absorbed and scattered Comptonized
  spectrum (model C; see \S\ref{sec:edge}).  Top panel: combined
  spectrum, with one \meg channel per bin. Second panel: spectrum
  rebinned to a S/N$\ge 5$ and $\ge4$ \meg channels per bin,
  \emph{without} refitting the spectrum. Third/fourth panels: Cash
  statistic residuals for the spectral binnings and fit shown in the
  first two panels.}
 \label{fig:chandra_spectra}
\end{figure}

We show the fit for model C in Figure~\ref{fig:chandra_spectra}.  The
inclusion of scattering and solid state absorption effects due to dust
grains reduces the overall required column to a value of $\nh \approx
(2.9\pm0.5) \times 10^{22}\,\cmt$.  The presence of a near edge
absorption feature is still required.  We note that in terms of
equivalent width all of the models discussed above have comparable
values, despite obvious changes in the absolute line depth as seen in
Figure~\ref{fig:edge_phenom}.  This is because the equivalent width is
a relative measure, and what is being deemed as ``continuum'' in the
equivalent width calculations includes the edge from the
absorption/scattering models.

\begin{deluxetable*}{cccccccccc}
\setlength{\tabcolsep}{0.03in} \tabletypesize{\footnotesize}
\tablewidth{0pt} \tablecaption{MCMC Line Significances}
\tablehead{\colhead{abs} & \colhead{Si K$\alpha$} & \colhead{near edge}
& \colhead{\ion{Si}{13}a} & \colhead{\ion{Si}{13}b} & \colhead{\ion{Si}{13}g} 
& \colhead{Ca K$\alpha$} & \colhead{\ion{Ca}{19}i} & \colhead{\ion{Ca}{19}r} 
& \colhead{\ion{Ca}{20}a}}
\startdata 
96.9\% & 66.3\% & 95.4\% 
& 99.999\% & 64.3\% & 95.9\% 
& 99.4\% & 99.1\% & 99.7\% 
& 97.0\% \\
\enddata 
\tablecomments{Significances from a Markov Chain Monte Carlo (MCMC)
  analysis of model C, subject to the line constraints discussed in
  the text.  Labels are the same as in Table~\ref{tab:pars}.
  Percentages are the fraction of the posterior probability
  that is $<0$ for absorption lines and $>0$ for emission
  lines. Significances do not include multiplicity of
  trials.}\label{tab:line_sig}
\end{deluxetable*}

We use model C, which has the most freedom to fit the Si edge region
with the neutral absorption and dust scattering models, to assess the
significances of the lines beyond the nominal 90\% confidence
intervals presented in Table~\ref{tab:pars}. We use this model in a
Markov Chain Monte Carlo (MCMC) calculation implemented in \isis
following the prescription of \cite{goodman:10a}.  (See our detailed
descriptions in \citealt{murphy:14a}.)  We evolve a set of 320
``walkers'' (ten initial models per free parameter, with their initial
parameters randomly distributed over the central 3\% of the 90\%
confidence intervals) for 40\,000 steps, of which we only retain the
last 2/3 for assessing probabilities (yielding over 8.5 million
samples in our posterior probability distributions).  The line widths
and energies are constrained as discussed above.
  
We calculate line significances as the fraction of the posterior
probability distribution with negative line amplitudes for absorption
lines, or the fraction of the posterior probability distribution with
positive amplitudes for emission lines.  This is of course a somewhat
local and constrained probability distribution that does not account
for any ``multiplicity of trials'' in our initial assessment of lines
to include in our models.  We present these line significances in
Table~\ref{tab:line_sig}.  In general, these significances are
commensurate with the results of the 90\% confidence intervals
presented in Table~\ref{tab:pars}, with the blueshifted \ion{Si}{13}
resonance line being the most significant feature.  The Si\,K$\alpha$
line is less significant than one might expect from
Table~\ref{tab:pars} owing to the fact that if the line energy shifts
from the best fit value by more than $\approx 5$\,eV in either
direction, a broader weak emission feature is allowed.

\begin{figure}
\begin{center}
\includegraphics[width=0.47\textwidth,viewport=90 40 579 520,clip]{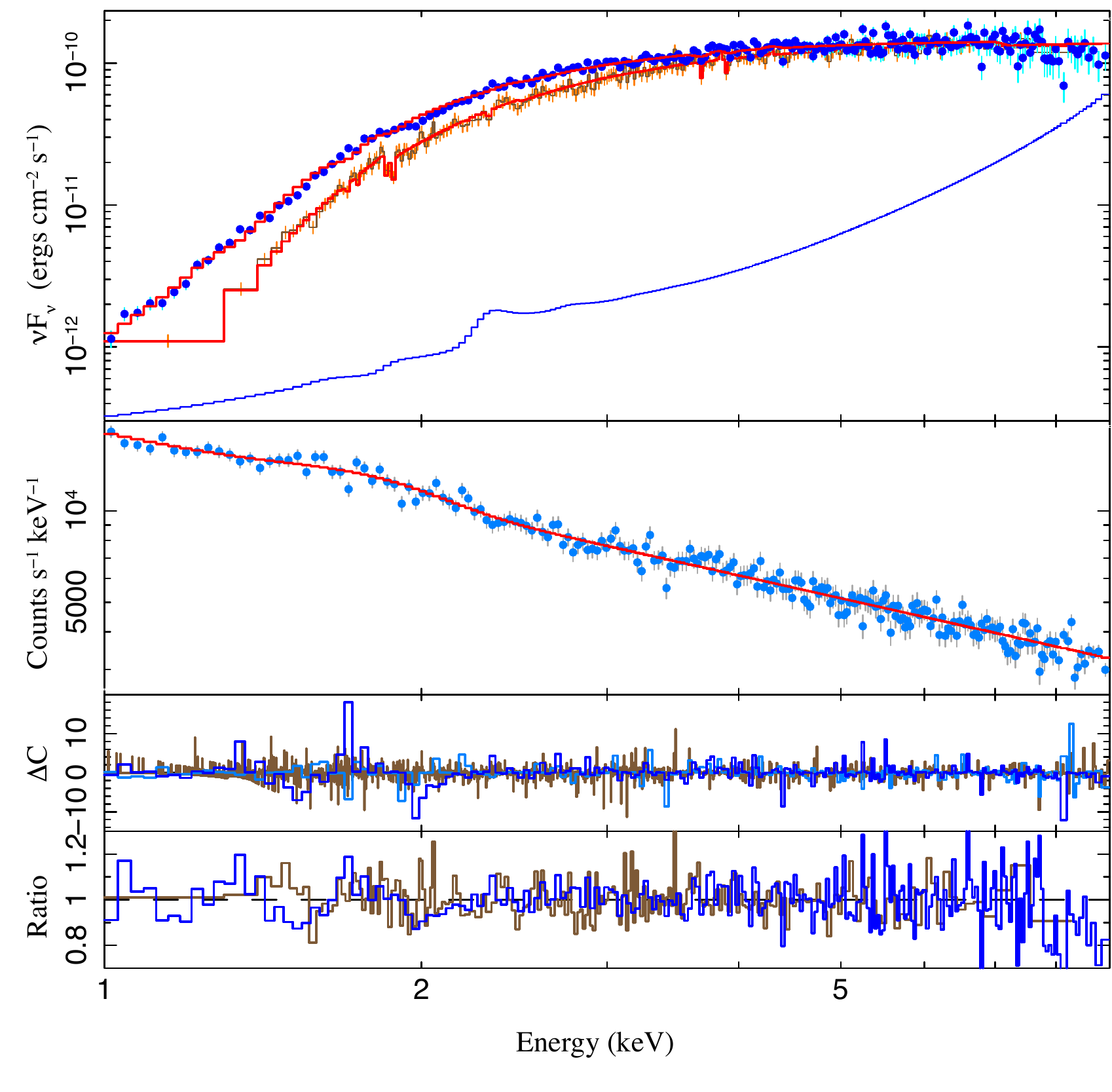}
\end{center}
\caption{Top panel: The flux-corrected combined \heg and \meg first
  order \chandra-\hetg spectra (brown histogram) and \nicer spectra
  (blue circles, pale blue histogram is the background model), jointly
  fit with an absorbed and scattered blackbody plus Comptonization
  model (model E). Owing to the different fields of view of each
  instrument, the dust scattering component is different in the
  \chandra and \nicer spectra, and accounts for the deviation between
  the two spectra at energies $\aproxlt 3$\,keV.  The \nicer spectra
  also have been renormalized to account for a fitted cross
  normalization constant between the \chandra-\hetgs and \nicer
  spectra. The second panel shows the simultaneous fit to the
  background spectra (see text).  The third and fourth panels show the
  residuals for the model fit.  The third panels shows the Cash
  statistics residuals for the \chandra-\hetgs (brown histogram) and
  \nicer source (blue histogram) and background (light blue histogram)
  spectra at the binning of the fit (see text).  The fourth panel
  shows the the data/model ratio residuals, but now omitting the
  \nicer background spectra and with the \chandra-\hetgs spectra
  rebinned for clarity, but without refitting the model.}
 \label{fig:joint_spectra}
\end{figure}

\subsection{Joint Fits with \nicer Spectra}\label{sec:nicer}

We next consider models C and D, but with the inclusion of 1--9\,keV
\nicer spectra. The \nicer instrument has a field of view $\approx
30$\,square arcmin, i.e., an $\approx 3$\arcmin\ radius
\citep{gendreau:16a}. We therefore expect a large fraction of the dust
scattered photons, lost from the \chandra-\hetgs spectra, to scatter
back into the field of view of \nicer (see the discussion of
\citealt{corrales:16a}).  Although these scattered photons are
time-delayed \citep{mccray:84a}, there is no indication that the
spectrum from tens of thousands of seconds earlier was substantially
different from what we observed.  As expected, fitting for the size of
the dust scattering halo relative to the \nicer PSF, we find
$\mathrm{H_{size}<1.9}$.  (We set the lower bound of
$\mathrm{H_{size}=0.01}$.)  That is, the spectra are consistent with a
substantial fraction (nearly all) of the scattered X-rays returning to
the \nicer field of view.  This is in fact apparent when comparing the
flux-corrected spectra between \chandra-\hetgs and \nicer, as seen in
Figure~\ref{fig:joint_spectra}.

\input{par_table.tex}

There are, however, significant residuals for the \nicer spectra in
the $\approx 1.5$--2.5\,keV region.  It is likely that both the fitted
equivalent neutral column, as well as the fraction of mass in dust ---
and specifically the fraction of mass in silicate dust --- is being
partly driven by the systematic uncertainties in the \nicer response
functions.  We have used MCMC analyses identically as described above
for all of our model fits to determine the interdependencies of the
fitted parameters, and to make confidence contours of these parameter
correlations. Although the contours of equivalent neutral column
vs.\ silicate dust mass column (Figure~\ref{fig:cross_norm}, left) are
consistent between \chandra-\hetgs and \nicer, the small \nicer
spectra error bars, coupled with large fit ratio residuals, indicate
that \nicer systematics in this regime are still a significant concern
for this aspect of the model fits.

To further bring agreement between the \chandra-\hetgs and \nicer
spectra, we have to include a cross-normalization between the two
detectors.  We choose an energy-independent cross-normalization
constant, with the only energy-dependent differences between the two
observations being the above expected differences due to the dust
scattering halo.  Although such energy-dependent cross calibration
differences may exist, we do not believe these data are sufficiently
constraining so as to allow exploration of a more complicated model.
For Model E, which allows the greatest freedom in the dust absorption
and scattering parameters, we also find the largest
cross-normalization constant, $0.83\pm0.02$.  This is somewhat lower
than one might initially expect from the 1--9\,keV \nicer flux, which
is $1.48\times 10^{-10}\,\ecs$ (see Figure~\ref{fig:lightcurve}).
Note, however, that the \nicer spectra are \emph{less affected} by the
dust halo, and therefore would have a slightly higher flux than
\chandra-\hetgs even if the cross normalization constant were unity.

In Figure~\ref{fig:cross_norm} we show the dependence of this cross
normalization constant on the fitted and/or presumed equivalent
neutral and silicate dust mass columns.  There are significant
systematic dependencies upon the latter, which is not surprising given
the ratio residuals present in Figure~\ref{fig:joint_spectra}.  For
the models that we have explored, however, we have not found a cross
normalization constant $\aproxgt 0.85$.

\section{Discussion}\label{sec:discuss}

We have presented a series of fits to \integral, \nicer, and
\chandra-\hetgs spectra of the AMXP \igr together with on-source NIR
observations performed within our collaboration.

\subsection{Comparison with previous findings} \label{sec:Dprevious}

Our IBIS/ISGRI spectrum of the brightest part of the hard X--ray
outburst of \igr~(24\,ks, Fig.~\ref{fig:integral_lcr}) results in a
non-attenuated power-law model ($\Gamma=2.0 \pm 0.2$) with no cut-off
required, and source detection up to about 110\,keV. This is
compatible with what was found in the dedicated \integral Target of
Opportunity observations of the source \citep[164\,ks;][]{kuiper:18a}
that significantly detected \igr up to 150\,keV using a powerlaw model
description. Such a high energy spectrum is representative for AMXPs
that are known to have quite high Comptonizing plasma temperatures ---
of the order of several tens of keV \citep[e.g.,][and references
  therein]{falanga:13a}, similar to the so-called Atoll LMXBs known to
host NS.

In agreement with the many preliminary analyses presented in the
references of \S\ref{sec:intro}, and specifically with the
\swift/\nicer/\nustar/\integral analyses presented by
\citet{sanna:18a}, we find that a highly absorbed powerlaw ($\nh
\approx 4\times10^{22}\,\cmt$, $\Gamma \approx 2$) describes the
spectra well.  When specifically modeling the 1--9\,keV spectra with a
Comptonized blackbody, fixing the coronal temperature to the 22\,keV
employed by \citet{sanna:18a}, there is an implied spectral curvature
in the 50--150\,keV \integral band that we do not detect. However, our
1--9\,keV spectra are largely insensitive to the temperature of the
corona, and instead are predominantly sensitive to the spectral slope
of Compton continuum which is $\Gamma \approx 2$ for all the models
that we have considered.

Our one major difference from the models of \citet{sanna:18a} is that
we fit a lower temperature, and hence a larger normalization, for the
seed photons input to Compton corona.  They found a blackbody emission
area consistent with the surface of a neutron star.  In contrast, the
blackbody normalizations presented in Table~\ref{tab:pars} imply
emission radii ranging from $\approx 1\,000$--$200\,000$\,km, if the
source is at the 8\,kpc distance of the Galactic bulge.  This would
imply that the seed photons for Comptonization were instead generated
by the accretion flow onto the neutron star, rather than its surface.

A second difference between our model fit results and previous fit
results using spectra from detectors with lower spectral resolution
than for \chandra-\hetgs concerns the fitted equivalent neutral
column.  Our model fits to $\nh$ are not only driven by the curvature
of the soft X-ray continuum spectra, but are also driven by direct
modeling of X-ray absorption edges of various atomic species.  The
advent of the era of high resolution spectroscopy is among the factors
that drove the development of the \texttt{tbvarabs} model
\citep{wilms:00a}.  This model utilizes improved knowledge of ISM
abundances and atomic cross sections, and it also provides a more
precise description of atomic edges from such species as O, Fe (via
the L and K edges), Ne, and for the case of \igr, the Si edge.
However, as pointed out by \citet{schulz:16a}, the \texttt{tbvarabs}
model under predicts the depth of the Si edge for a given equivalent
neutral column.  Phenomenologically, this can be accounted for by
either adding a separate Si-region edge to the model (while
artificially reducing the Si abundance) with $\mathrm{\tau_{edge} =
  0.19\pm0.05}$, or by increasing the Si abundance in the model to
$\mathrm{A_{Si} = 1.76^{+0.48}_{-0.47}}$ (models A and B in
Table~\ref{tab:pars}).  This is in complete agreement with the results
of \citet{schulz:16a} and also with newer abundance measurements for
B-stars in the Galaxy \citep{nieva:12a}, which imply
$\mathrm{A_{Si}}=1.70$.

\begin{figure*}
\begin{center}
\includegraphics[width=0.32\textwidth,viewport=90 20 579 382,clip]{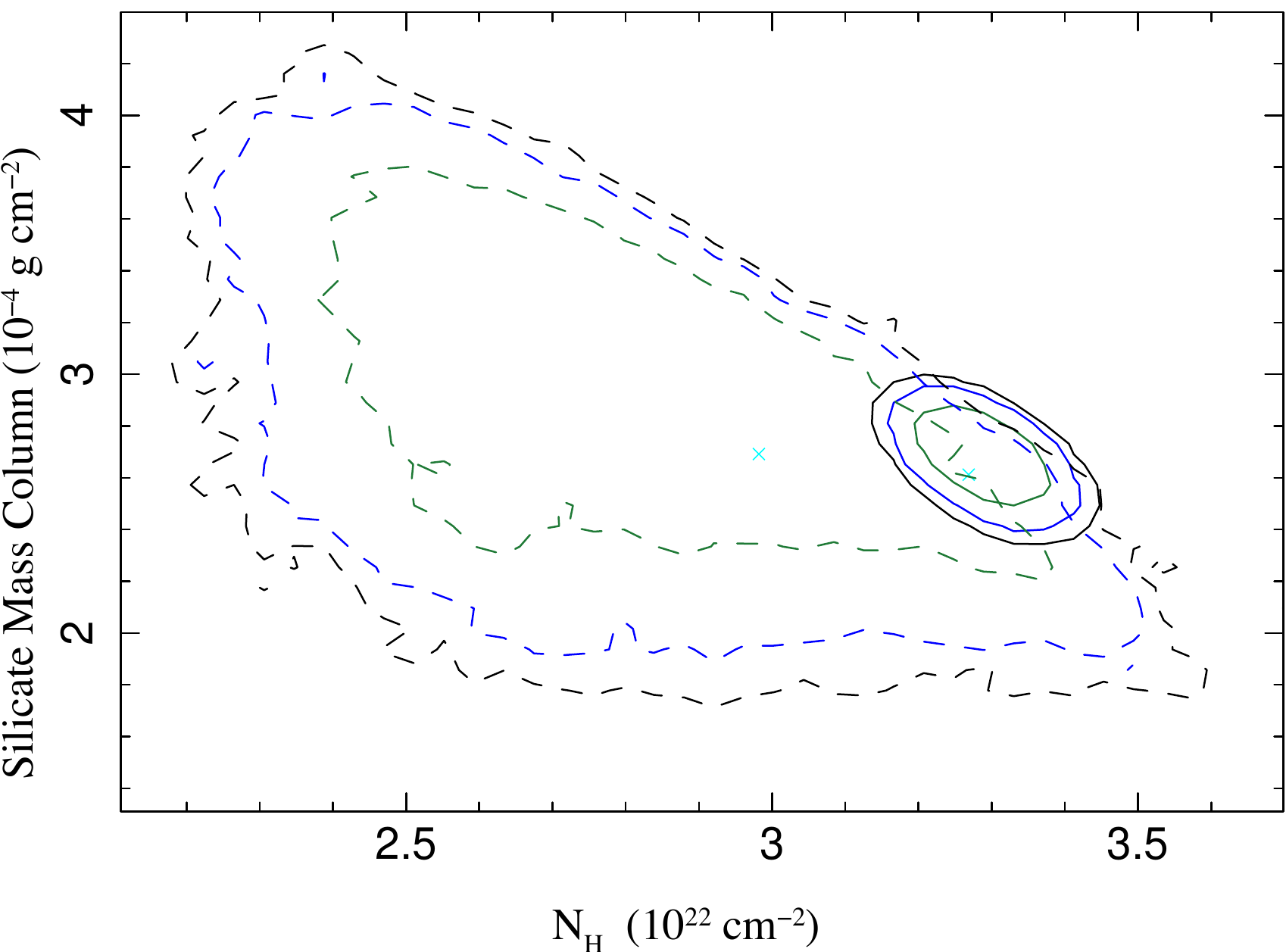}
\includegraphics[width=0.32\textwidth,viewport=90 20 579 382,clip]{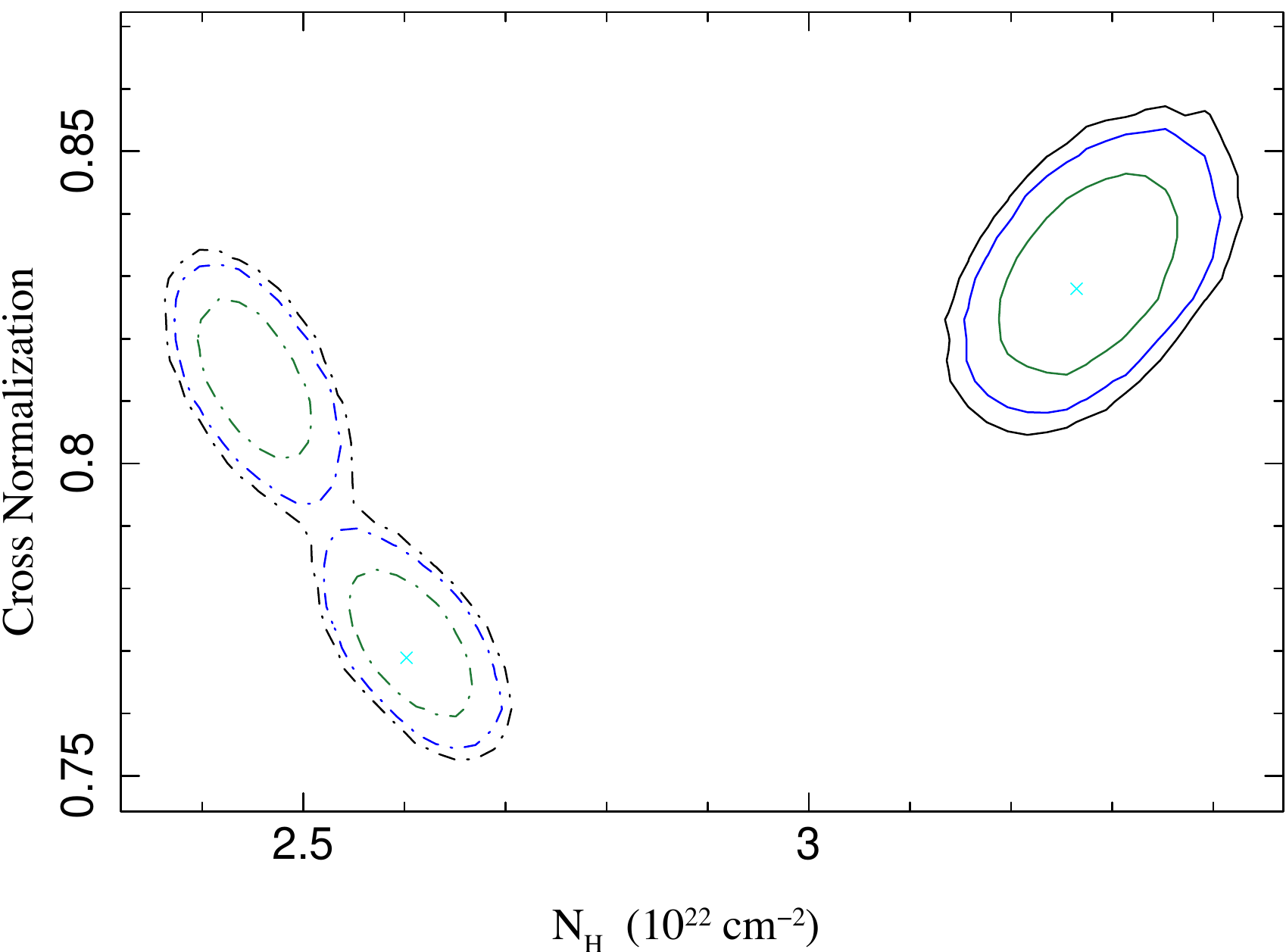}
\includegraphics[width=0.32\textwidth,viewport=90 20 579 382,clip]{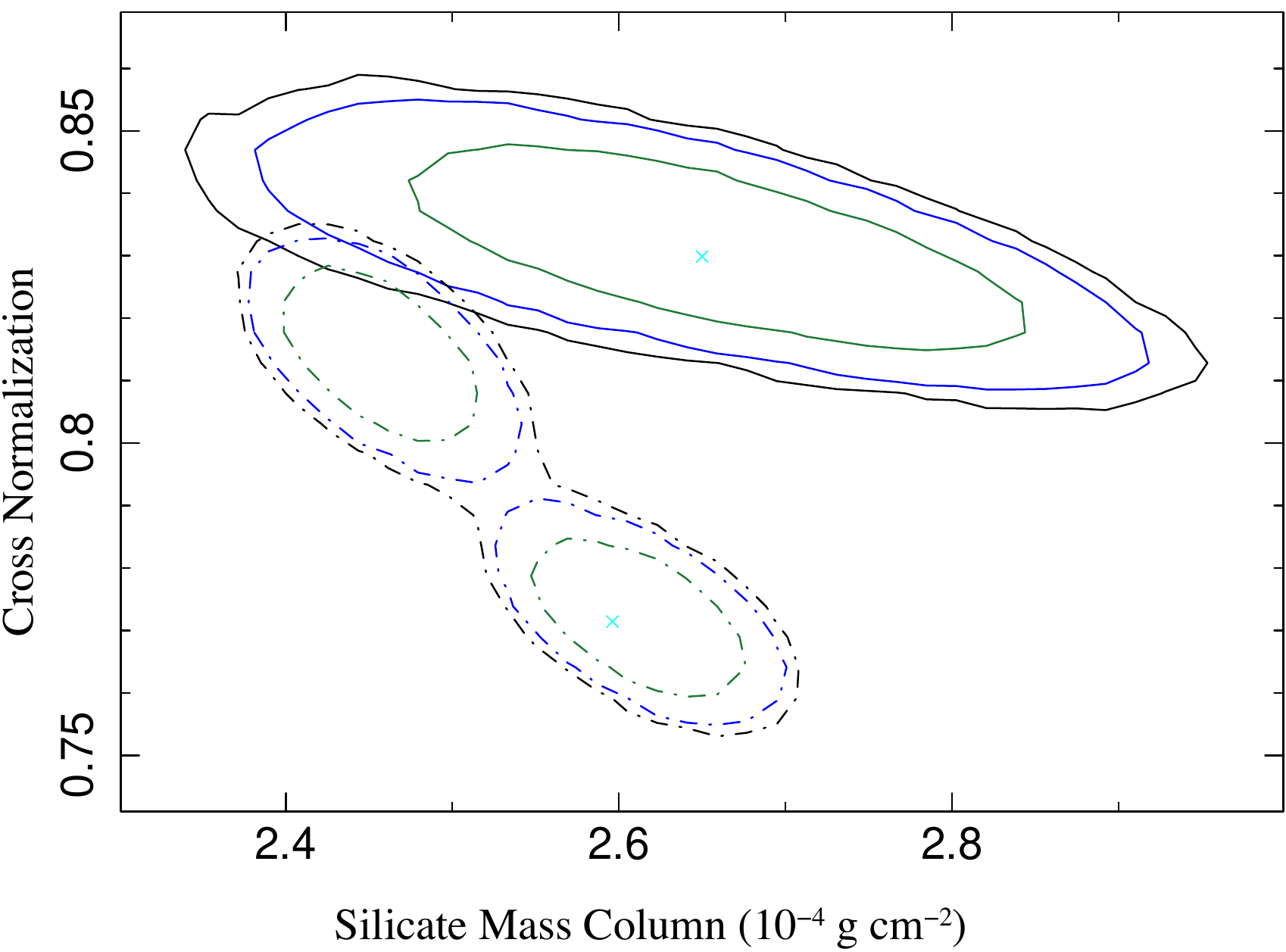}
\end{center}
\caption{Contours for interstellar absorption and dust absorption and
  scattering parameters for absorbed/scattered blackbody plus
  Comptonization fits to the 1--9\,keV \chandra-\hetg spectra on their
  own (dashed contours--- model C), or in combination with the
  1--9\,keV \nicer spectrum (solid--- model E --- and dash-dot---
  model F --- contours), as derived from Markov Chain Monte Carlo
  (MCMC) analyses of the model fits. \textsl{Left:} Equivalent
  hydrogen column for gas absorption vs.\ silicate mass column for
  dust scattering and absorption.  \textsl{Middle:} Cross
  normalization value for the \nicer spectrum (relative to
  \chandra-\hetg normalization) vs.\ equivalent hydrogen column.
  Solid contours are for a freely variable dust mass and silicate
  fraction (see \S\ref{sec:edge}), while the dash-dot contours presume
  a fixed dust mass fraction of 0.01 and a fixed silicate fraction of
  0.6.  \textsl{Right:} \nicer cross normalization constant
  vs.\ silicate mass column for both freely variable and fixed dust
  mass and silicate fractions.}
 \label{fig:cross_norm}
\end{figure*}

\subsection{Dust absorption/scattering and source distance} \label{sec:Ddust}
A more physical description of this result is provided by employing
the models of \citet{corrales:16a}.  As pointed out by these authors,
as an absorption model the \texttt{tbvarabs} model does not account
for soft X-ray scattering or solid state absorption effects by dust
except for shielding.  Both are important for the high \emph{spatial}
resolution observations of \chandra-\hetgs.  Dust scatters X-rays out
of our line of site on arcsec size scales, but it scatters back into
the line of site, albeit with a time delay, on arcminute scales
\citep{mccray:84a}.  Thus, we have to account for both dust scattering
and solid state absorption effects in modeling the \chandra-\hetgs
spectra of \igr.  We have done this in models C and D from
Table~\ref{tab:pars}, using the dust models of \citet{corrales:16a}.
The inclusion of dust has the effect of \emph{reducing} the required
equivalent neutral column.  Since it contains the least restrictive
assumptions about the column mass fraction in dust or the fraction of
dust in silicates, we consider model C, with $\nh =
(2.9\pm0.5)\times10^{22}\,\cmt$, to be our most fair estimate for the
equivalent neutral column\footnote{Again using the \citet{predehl:95a}
  and \citet{fitzpatrick:99a} relationships between extinction and
  equivalent column, this implies I and K band absorptions of
  $\mathrm{A_V} = 16.2$\,mag and $\mathrm{A_K} = 1.85$\,mag. Both
  absorptions are high, and are still consistent with a non-detection
  in the I-band, as discussed in \S\ref{sec:multi}.}  along our line
of site to \igr.

This value is lower, by $\approx 1/5$--1/3, compared to model
estimates made without accounting for dust effects (e.g.,
\citealt{sanna:18a,russell:18a}).  As discussed by
\citet{russell:18a}, the equivalent neutral column inferred from
reddening maps is $\nh < 2.2 \times 10^{22}\,\cmt$ (for the full
column along the line of site), and is $\nh > 0.7 \times
10^{22}\,\cmt$ for distances $>6$\,kpc.  Thus \citet{russell:18a}
argue for a large distance, at the Galactic bulge distance or beyond,
and further argue that \igr is radio bright for an AMXP.  (In fact,
based upon its radio brightness relative to its X-ray flux, \igr was
initially hypothesized to be a black hole candidate;
\citealt{russell:18c}.)  \citet{russell:18a} offer the alternative
hypothesis that if much of the absorption is local to the system, then
it can be significantly closer allowing for a more typical ratio of
radio to X-ray luminosity for an AMXP.  (Empirically, the radio flux
drops more slowly than X-ray flux for decreasing luminosities.)

Although our inclusion of the dust effects lowers the fitted
equivalent neutral column, it does not do so substantially enough to
fundamentally alter the conclusions\footnote{It should be noted that
  ``equivalent neutral column'' is often used as a proxy parameter;
  however, it is not always clear within the literature to what degree
  this parameter is \emph{the same} for different types of
  measurements.  That is, what are the systematic differences between
  this parameter when discussing X-ray absorption vs.\ X-ray dust
  halos vs.\ interstellar reddening vs.\ 21\,cm measurements?
  Discussing the potential systematic differences for the equivalent
  neutral column used in each type of such measurements is well beyond
  the scope of this work.  However, this does not alter the basic
  conclusion that our measured column would have to be predominantly
  local to the source in order to have \igr be substantially closer
  than the Galactic bulge distance.}  of \citet{russell:18a}.

A further argument in favor of the source being at a large distance
with a column primarily attributable to the ISM (as opposed to local
absorption) is the presence of substantial near edge absorption
feature at an energy of 1.848\,keV.  Such a near edge absorption
feature is routinely seen in X-ray binary sources with columns in the
range of $\approx(1$--$\mathrm{8)\times10^{22}\,cm^{-2}}$
\citep{schulz:16a}; however, the near edge feature is often variable
and of lower equivalent width magnitude than we observe here.  The
speculation is that dust local to the system is destroyed/ionized by
the source's X-rays.  \citet{schulz:16a} essentially fit a spectral
model equivalent to model A in Table~\ref{tab:pars}, and the highest
magnitude equivalent widths they find are $\approx -8\pm2$\,mA for
$\nh \aproxgt 4 \times 10^{22}\,\cmt$.  For \igr fit with model A, we
find $\mathrm{EW}=-4.1^{+2.0}_{-1.7}$\,eV $=-16^{+8}_{-7}$\,mA. If
this feature were primarily local and subject to destruction by
ionization due to the source, it would be unusual to find its
equivalent width at a magnitude greater than observed in the entire
\citet{schulz:16a} sample, while at the same time also seeing a
\ion{Si}{13} absorption line with a high magnitude equivalent width
($\mathrm{EW}=-4.3^{+0.7}_{-1.7}$\,eV, $=-15^{+3}_{-6}$\,mA for model
A) outflowing at 0.0093\,c.  Thus we hypothesize that a large fraction
of the observed column is associated with the ISM (as is consistent
with our NIR results discussed in \S\ref{sec:multi}), the source is at
a large distance, and hence its radio flux is indeed high for an AMXP.

\subsection{Outflowing wind} \label{sec:Doutflow}
The \ion{Si}{13} absorption line indicates a mass outflow in the \igr
system.  We can constrain the energy flux associated with this outflow
based upon the line equivalent width.  Assuming that the line is on
the linear part of the curve of growth, $\mathrm{W_\lambda}$, its
equivalent width in \AA, is related to the \ion{Si}{13} column,
$\mathrm{N_{Si13}}$, by
\begin{equation}
\mathrm{\frac{W_\lambda}{\lambda} = 
   \frac{\pi e^2}{m_e c^2}\,N_{Si13}\,\lambda\,f_{ij}
   = (8.85 \times 10^{-13}\,cm) \, N_{Si13}\,\lambda\,f_{ij} }
\end{equation}
\citep{spitzer:78a}. Using an oscillator strength of 0.75
\citep{kramida:18a}, the column is $N_\mathrm{Si13} = 5\times
  10^{16}\,\mathrm{cm}^{-2}$, which yields a wind kinetic energy flux of
\begin{equation}
E_\mathrm{wind} = 2 \times
  10^{13}\,\left(\frac{f_{\mathrm{Si13}}}{0.1}\right)^{-1}\,\mathrm{erg}\,\mathrm{cm}^{-2}\,\mathrm{s}^{-1}~~,
\end{equation}
where $f_\mathrm{Si13}$ is the fraction of Si in \ion{Si}{13}, and we
have used the ISM abundances of \citet{wilms:00a} in going from an Si
column to a hydrogen column.

In order to determine the total kinetic energy luminosity, we would
need to know the characteristic wind radius.  The large effective
radius of the blackbody seed photons suggests a large wind launching
radius, $\aproxgt 10^{10}$\,cm.  The narrow width of the \ion{Si}{13}
line suggests an even larger radius, $\aproxgt 10^{11}$\,cm.  The
kinetic energy luminosity of the wind then becomes
\begin{equation}
\mathrm{E_{wind} = 2 \times
  10^{35} \left(\frac{f_{Si13}}{0.1}\right)^{-1}\left( \frac{\Omega_{wind}}
  {4 \pi}\right)\left( \frac{R_{wind}}{10^{11}\,cm} \right )\,erg\,s^{-1}~,}
\end{equation}
where $\mathrm{\Omega_{wind}}$ is the solid angle subtended by the
wind and $\mathrm{R_{wind}}$ is the wind launching radius.  This is
potentially a large fraction of the luminous energy of the source. 

Of the additional lines that we included in our models, the only
plausible identifications that we have are with various species of Ca.
These are not at a consistent set of velocity shifts, nor even all in
emission or absorption.  If the line identifications are real, these
lines could be associated with a variety of locations in the accretion
flow and/or the atmosphere of the companion star, and may indicate an
overabundance of calcium in the system.  It is possible that the
progenitor of the \igr system was the collapse of a white dwarf,
producing a calcium rich Type Ib supernova (\citealt{perets:10a}; see
also \citealt{canal:90a,metzger:09a}); one possible example of such a
system comes from optical/\chandra observations of a NS binary system
with calcium overabundance of a factor of 6, within the supernova
remnant RCW 86, that likely will evolve into a LMXB system
\citep{gvaramadze:17a}.  \igr may be a later evolutionary stage of
such a system.  

Theoretical scenarios show a clear variety of evolutionary channels in
LMXBs and it is not easy to estimate the presence/amount of Ca
therein, especially when subject to a long-term (possibly
intermittent) X-ray irradiation that dramatically alters the evolution
of the system, be it by irradiation-driven winds and/or expansion of
the companion \citep[e.g.,][]{podsiadlowski2002, nelson2003,
  tauris2006}. However, highly ionized atmospheres or winds are known
to be present in LMXBs and are detected as warm emitters and/or
absorbers in many systems \citep[][and references
  therein]{diaz2016}. \ion{Ca}{20} absorption lines have been detected
in the \xmm spectra of GX~13$+$1 \citep{sidoli2002,ueda2004} as well
as in MXB~1659$-$298 \citep{ponti2018}. Similarly, the presence of Ca
has been observed in the AMXP SAX~J1748.9$-$2021 \citep{pintore2016}
as well as in the binary millisecond pulsar PSR~J1740$-$5340
\citep{sabbi2003}. These findings, together with our results on \igr,
seem to suggest that the accretion flow and/or companion atmosphere
can be Ca-rich if the companion is subject to prolonged mass loss and
interactions with the millisecond pulsar.

\subsection{A multi-facility approach: final considerations} \label{sec:Dfinal}

AMXPs are known to have X-ray spectra characterized by high
Comptonizing plasma temperatures, of the order of several tens of keV
\citep[e.g.,][and references therin]{falanga:13a}, similarly to the
so-called Atoll LMXBs known to host NS. This results in non-attenuated
power-law spectra up to hundreds of keV, compatible with what we found
for the brightest part of the outburst.

On the lower-energy part of the spectrum, we note that overall there
is good agreement between the \chandra-\hetgs and \nicer 1--9\,keV
spectra, if one carefully accounts for the manner in which each
instrument views the scattering by the dust halo in front of \igr. The
differences seen between the two flux-corrected spectra in
Figure~\ref{fig:joint_spectra} are primarily due to the effects of
dust scattering, rather than due to uncertainties in instrumental
response.  As regards the instrumental response, essentially all of
our detailed information regarding absorption and outflows in the \igr
system comes from the high resolution \hetgs.  \nicer lacks both the
spectral resolution, and currently has significant response
uncertainties, in the $\approx 2$\,keV region.  There also remains an
$\aproxgt 15\%$ normalization difference between the \chandra-\hetgs
and \nicer spectra (Figure~\ref{fig:cross_norm}).  On the other hand,
\chandra-\hetgs is incapable of achieving the time resolution of
\nicer that was required to characterize the pulsar and orbital
periods of the \igr system (see \citealt{sanna:18a}).

Together these instruments, along with the radio
and NIR measurements discussed above, paint a picture of \igr as a
somewhat distant system with a high velocity outflow and an unusually
bright radio flux for an AMXP, that might have formed from a rare,
calcium rich supernova explosion.

\acknowledgements The authors thank Lia Corrales for useful
discussions concerning dust scattering, and members of the \nicer
team, especially Paul Ray, for discussions concerning the \nicer
spectra.  Michael Nowak gratefully acknowledges funding support from
the National Aeronautics and Space Administration through Chandra
Guest Observer Grant GO8-19022X. Adamantia Paizis acknowledges
financial contribution from ASI/INAF n.2013-025.R1 contract and from
the agreement ASI-INAF n.2017-14-H.0. J\'er\^ome Rodriguez
acknowledges partial funding from the French Space Agency (CNES).
Sylvain Chaty and Francis Fortin are grateful to the Centre National
d'Etudes Spatiales (CNES) for the funding of MINE (Multi-wavelength
INTEGRAL Network).  Gaurava Jaisawal acknowledges support from the
Marie Sk{\l}odowska-Curie Actions grant no. 713683 (H2020;
COFUNDPostdocDTU).  Based on observations with \integral, an ESA
project with instruments and science data centre funded by ESA member
states (especially the PI countries: Denmark, France, Germany, Italy,
Spain, and Switzerland), Czech Republic and Poland, and with the
participation of Russia and the USA.  Based on observations collected
at the European Organisation for Astronomical Research in the Southern
Hemisphere under ESO programme(s) 0101.D-0082(A).

\facilities{CXO (ACIS, HETG), INTEGRAL, NICER, Swift (XRT), ESO, VLT
  (HAWK-I)}

\software{XSPEC, ISIS}


\end{document}

%% file: par_table.tex
\begin{center} 
\begin{deluxetable*}{rcrrrrrr} 
\setlength{\tabcolsep}{0.03in}\tabletypesize{\footnotesize} 
\tablewidth{0pt} \tablecaption{Spectral Fit Parameters} 
\tablehead{ \colhead{Parameter} & \colhead{Units} 
& \colhead{A} & \colhead{B} & \colhead{C} 
& \colhead{D} & \colhead{E} & \colhead{F} }
\startdata 
$\nh$ & $10^{22}\,\cmt$ & 
${4.23}^{+0.16}_{-0.14}$ & ${4.21}^{+0.18}_{-0.15}$ & ${2.86}^{+0.54}_{-0.55}$ &
${2.62}^{+0.07}_{-0.07}$ & ${3.28}^{+0.10}_{-0.09}$ & ${2.63}^{+0.05}_{-0.21}$  \\  
$\mathrm{A_{Si}}$ &  & 
\nodata & ${1.76}^{+0.48}_{-0.47}$ & \nodata &
\nodata & \nodata & \nodata  \\  
$\mathrm{E_{edge}}$ & keV & 
${1.859}^{+0.004}_{-0.020}$ & \nodata & \nodata &
\nodata & \nodata & \nodata  \\  
$\mathrm{\tau_{edge}}$ &  & 
${0.19}^{+0.05}_{-0.05}$ & \nodata & \nodata &
\nodata & \nodata & \nodata  \\  
$\mathrm{f_{dust}}$ & $10^{-2}$ & 
\nodata & \nodata & ${0.71}^{+0.60}_{-0.26}$ &
\textsl{1} & ${0.49}^{+0.04}_{-0.04}$ & \textsl{1}  \\  
$\mathrm{f_{silicate}}$ &  & 
\nodata & \nodata & ${0.92}^{+0.08}_{-0.49}$ &
\textsl{0.6} & ${1.0}_{-0.06}$ & \textsl{0.6}  \\  
$\mathrm{C_n}$ &  & 
\nodata & \nodata & \nodata &
\nodata & ${0.83}^{+0.02}_{-0.02}$ & ${0.77}^{+0.01}_{-0.01}$  \\  
$\mathrm{H_{size}}$ &  & 
\nodata & \nodata & \nodata &
\nodata & ${0.01}^{+1.33}_{-0.0}$ & ${1.85}^{+0.06}_{-0.06}$  \\  
$\mathrm{N_{nc}}$ &  & 
${0.09}^{+0.01}_{-0.01}$ & ${0.09}^{+0.01}_{-0.01}$ & ${0.10}^{+0.02}_{-0.01}$ &
${0.10}^{+0.01}_{-0.01}$ & ${0.10}^{+0.01}_{-0.01}$ & ${0.09}^{+0.01}_{-0.01}$  \\  
$\mathrm{\Gamma_{nc}}$ &  & 
${1.98}^{+0.06}_{-0.07}$ & ${1.99}^{+0.07}_{-0.07}$ & ${2.06}^{+0.08}_{-0.07}$ &
${2.05}^{+0.05}_{-0.05}$ & ${2.02}^{+0.04}_{-0.02}$ & ${1.96}^{+0.03}_{-0.03}$  \\  
$\mathrm{kT_{nc}}$ & keV & 
${0.084}^{+0.003}_{-0.003}$ & ${0.082}^{+0.003}_{-0.002}$ & \textsl{0.06} &
\textsl{0.06} & ${0.057}^{+0.001}_{-0.001}$ & ${0.074}^{+0.002}_{-0.007}$  \\  
$\mathrm{N_{bb}}$ & $10^7$ & 
${5.5}^{+1230}_{-5.2}$ & ${8.1}^{+630}_{-7.7}$ & ${330}^{+470}_{-270}$ &
${400}^{+390}_{-290}$ & ${2000}^{+3800}_{-1900}$ & ${30.7}^{+230}_{-22.2}$  \\  
$\mathrm{E_{abs}}$ & keV & 
${1.6949}^{+0.0005}_{-0.0008}$ & ${1.6949}^{+0.0006}_{-0.0008}$ & ${1.6949}^{+0.0005}_{-0.0008}$ &
${1.6949}^{+0.0006}_{-0.0008}$ & \textsl{1.6949} & \textsl{1.6949}  \\  
$\mathrm{\sigma_{abs}}$ & eV & 
${0.2}^{+0.9}_{-0.1}$ & ${0.2}^{+0.9}_{-0.1}$ & ${0.2}^{+0.9}_{-0.1}$ &
${0.2}^{+0.9}_{-0.1}$ & \textsl{0.2} & \textsl{0.2}  \\  
$\mathrm{EW_{abs}}$ & eV & 
${-2.3}^{+1.1}_{-0.8}$ & ${-2.3}^{+0.9}_{-0.8}$ & ${-2.3}^{+0.6}_{-1.2}$ &
${-2.3}^{+0.9}_{-0.8}$ & \textsl{-2.3} & \textsl{-2.3}  \\  
$\mathrm{E_{Si\,K\alpha}}$ & keV & 
${1.735}^{+0.001}_{-0.001}$ & ${1.735}^{+0.001}_{-0.001}$ & ${1.735}^{+0.001}_{-0.001}$ &
${1.735}^{+0.001}_{-0.001}$ & \textsl{1.735} & \textsl{1.735}  \\  
$\mathrm{\sigma_{Si\,K\alpha}}$ & eV & 
${0.2}^{+1.3}_{-0.1}$ & ${0.3}^{+1.1}_{-0.2}$ & ${0.2}^{+1.3}_{-0.1}$ &
${0.2}^{+1.3}_{-0.1}$ & \textsl{0.2} & \textsl{0.2}  \\  
$\mathrm{EW_{Si\,K\alpha}}$ & eV & 
${-2.3}^{+1.1}_{-0.8}$ & ${-2.3}^{+1.2}_{-0.8}$ & ${-2.2}^{+0.4}_{-0.9}$ &
${-2.3}^{+1.2}_{-0.8}$ & \textsl{-2.2} & \textsl{-2.3}  \\  
$\mathrm{E_{near\,edge}}$ & keV & 
${1.848}^{+0.002}_{-0.001}$ & ${1.848}^{+0.002}_{-0.002}$ & ${1.848}^{+0.002}_{-0.003}$ &
${1.848}^{+0.002}_{-0.002}$ & \textsl{1.848} & \textsl{1.848}  \\  
$\mathrm{\sigma_{near\,edge}}$ & eV & 
${2.1}^{+2.1}_{-2.0}$ & ${1.6}^{+2.6}_{-1.5}$ & ${1.8}^{+3.0}_{-1.7}$ &
${1.8}^{+2.9}_{-1.7}$ & \textsl{1.8} & \textsl{1.8}  \\  
$\mathrm{EW_{near\,edge}}$ & eV & 
${-4.1}^{+2.0}_{-1.7}$ & ${-3.3}^{+1.5}_{-2.0}$ & ${-3.3}^{+2.2}_{-0.5}$ &
${-3.3}^{+2.4}_{-0.1}$ & \textsl{-3.3} & \textsl{-3.3}  \\  
$\mathrm{z_{Si\,XIII}}$ &  & 
${0.0093}^{+0.0006}_{-0.0009}$ & ${0.0093}^{+0.0006}_{-0.0010}$ & ${0.0092}^{+0.0007}_{-0.0008}$ &
${0.0092}^{+0.0006}_{-0.0008}$ & \textsl{0.0092} & \textsl{0.0092}  \\  
$\mathrm{\sigma_{Si\,XIII}}$ & eV & 
${1.5}^{+2.2}_{-1.2}$ & ${1.5}^{+2.3}_{-1.2}$ & ${1.7}^{+1.8}_{-1.4}$ &
${1.8}^{+1.9}_{-1.5}$ & \textsl{1.8} & \textsl{1.8}  \\  
$\mathrm{EW_{Si\,XIIIa}}$ & eV & 
${-4.3}^{+0.7}_{-1.7}$ & ${-4.4}^{+1.5}_{-1.8}$ & ${-4.7}^{+1.8}_{-1.5}$ &
${-4.7}^{+1.8}_{-1.6}$ & \textsl{-4.7} & \textsl{-4.7}  \\  
$\mathrm{EW_{Si\,XIIIb}}$ & eV & 
${-0.7}^{+2.9}_{-2.1}$ & ${-0.6}^{+2.8}_{-2.1}$ & ${-0.5}^{+2.9}_{-2.0}$ &
${-0.5}^{+2.9}_{-2.2}$ & \textsl{-0.5} & \textsl{-0.5}  \\  
$\mathrm{EW_{Si\,XIIIg}}$ & eV & 
${-2.3}^{+2.9}_{-3.3}$ & ${-2.3}^{+2.9}_{-3.3}$ & ${-2.4}^{+3.1}_{-3.0}$ &
${-2.4}^{+2.6}_{-3.1}$ & \textsl{-2.4} & \textsl{-2.4}  \\  
$\mathrm{E_{Ca\,K\alpha}}$ & keV & 
${3.687}^{+0.007}_{-0.006}$ & ${3.687}^{+0.007}_{-0.007}$ & ${3.687}^{+0.007}_{-0.006}$ &
${3.687}^{+0.007}_{-0.006}$ & \textsl{3.687} & \textsl{3.687}  \\  
$\mathrm{\sigma_{Ca\,K\alpha}}$ & eV & 
${4.3}^{+7.8}_{-4.2}$ & ${4.3}^{+7.8}_{-4.2}$ & ${4.4}^{+7.8}_{-4.3}$ &
${4.3}^{+7.8}_{-4.2}$ & \textsl{4.4} & \textsl{4.3}  \\  
$\mathrm{EW_{Ca\,K\alpha}}$ & eV & 
${-6.1}^{+3.2}_{-3.2}$ & ${-6.1}^{+3.2}_{-3.2}$ & ${-6.2}^{+3.8}_{-3.3}$ &
${-6.1}^{+3.2}_{-3.2}$ & \textsl{-6.2} & \textsl{-6.1}  \\  
$\mathrm{E_{Ca\,XIXi}}$ & keV & 
${3.845}^{+0.028}_{-0.014}$ & ${3.845}^{+0.027}_{-0.014}$ & ${3.845}^{+0.029}_{-0.015}$ &
${3.845}^{+0.027}_{-0.014}$ & \textsl{3.845} & \textsl{3.845}  \\  
$\mathrm{\sigma_{Ca\,XIXi}}$ & eV & 
${0.7}^{+19.3}_{-0.6}$ & ${1.1}^{+18.9}_{-1.0}$ & ${1.0}^{+19.0}_{-0.9}$ &
${0.6}^{+19.4}_{-0.5}$ & \textsl{1.0} & \textsl{0.6}  \\  
$\mathrm{EW_{Ca\,XIXi}}$ & eV & 
${6.1}^{+15.8}_{-3.5}$ & ${6.1}^{+15.5}_{-3.5}$ & ${6.0}^{+61.6}_{-3.6}$ &
${6.1}^{+29.2}_{-3.6}$ & \textsl{6.0} & \textsl{6.1}  \\  
$\mathrm{E_{Ca\,XIXr}}$ & keV & 
${3.896}^{+0.011}_{-0.008}$ & ${3.896}^{+0.011}_{-0.008}$ & ${3.896}^{+0.011}_{-0.008}$ &
${3.896}^{+0.011}_{-0.008}$ & \textsl{3.896} & \textsl{3.896}  \\  
$\mathrm{\sigma_{Ca\,XIXr}}$ & eV & 
${6.4}^{+12.8}_{-6.3}$ & ${6.4}^{+12.9}_{-6.3}$ & ${6.5}^{+13.0}_{-6.4}$ &
${6.4}^{+12.9}_{-6.3}$ & \textsl{6.5} & \textsl{6.4}  \\  
$\mathrm{EW_{Ca\,XIXr}}$ & eV & 
${-6.6}^{+3.9}_{-4.4}$ & ${-6.6}^{+3.9}_{-4.5}$ & ${-6.7}^{+4.0}_{-4.4}$ &
${-6.6}^{+3.9}_{-4.2}$ & \textsl{-6.7} & \textsl{-6.6}  \\  
$\mathrm{E_{Ca\,XXa}}$ & keV & 
${4.295}^{+0.007}_{-0.010}$ & ${4.295}^{+0.006}_{-0.009}$ & ${4.294}^{+0.007}_{-0.006}$ &
${4.296}^{+0.005}_{-0.011}$ & \textsl{4.294} & \textsl{4.294}  \\  
$\mathrm{\sigma_{Ca\,XXa}}$ & eV & 
${0.8}^{+1.3}_{-0.7}$ & ${0.2}^{+0.0}_{-0.1}$ & ${0.3}^{+2.7}_{-0.2}$ &
${0.3}^{+0.2}_{-0.2}$ & \textsl{0.5} & \textsl{0.3}  \\  
$\mathrm{EW_{Ca\,XXa}}$ & eV & 
${7.0}^{+5.3}_{-4.3}$ & ${6.9}^{+5.2}_{-4.2}$ & ${6.9}^{+1.1}_{-4.5}$ &
${6.8}^{+0.9}_{-4.2}$ & \textsl{6.9} & \textsl{6.8}  \\  
Cash/DoF & & 1958.7/2170  & 1959.6/2171  & 1956.6/2171  & 1957.8/2173  & 2507.0/2645  & 2597.1/2647  \\ 
\enddata \tablecomments{ All errors are 90\% confidence level for one
  interesting parameter (determined as $\Delta$Cash=2.71, which
    is correct in the limit that Cash statistics approach $\chi^2$
    statistics).  Models A--D are for \chandra-\hetg 1-9\,keV spectra
  only, while models E and F also include \nicer 1-9\,keV spectra.
  Italicized parameters were held fixed at that value. See text for a
  description of the models and model parameters. }\label{tab:pars}
 \end{deluxetable*} 
\end{center}